\documentclass[seceq]{ptptex}

\usepackage{bm}
\usepackage{graphicx}
\usepackage{mathrsfs} 
\usepackage{cases}

\newcommand{\bx}{{\bm{x}}}
\newcommand{\br}{{\bm{r}}}
\newcommand{\bk}{{\bm{k}}}
\newcommand{\bp}{{\bm{p}}}
\newcommand{\bq}{{\bm{q}}}

\newcommand{\integ}{\int\!\!}

\newcommand{\out}{_{\rm out}}
\newcommand{\side}{_{\rm side}}
\newcommand{\lon}{_{\rm long}}
\title{%        %You can use \\ for explicit line-break
HBT Pion Interferometry 

with Phenomenological Mean Field Interaction
}

\author{%       %Use \scshape  for the family name
Koichi \textsc{Hattori}$^{1,2}$\footnote{E-mail: khattori@post.kek.jp} 
}

\inst{%     %Affiliation, neglected when [addenda] or [errata]
$^1$Institute of Physics, University of Tokyo, Tokyo 153-8902, Japan 

\hspace{-1.7cm}$^2$High Energy Accelerator Research Organization (KEK), 

\hspace{-2cm}Tsukuba, 305-0801, Japan\footnote{Current affiliation.} 
}

\abst{%         %this abstract is neglected when [addenda] or [errata]
To extract information on hadron production dynamics 
in the ultrarelativistic heavy ion collision, 
the space-time structure of the hadron source has been measured 
using Hanbury Brown and Twiss interferometry. 
We study the distortion of the source images due to the effect of a final state interaction. 
We describe the interaction, taking place 
during penetrating through a cloud formed by evaporating particles, 
in terms of a one-body mean field potential localized in the vicinity of the source region. 
By adopting the semiclassical method, 
the modification of the propagation of an emitted particle is examined. 
In analogy to the optical model applied to nuclear reactions, 
our phenomenological model has an imaginary part of the potential, 
which describes the absorption in the cloud. 
In this work, we focus on the pion interferometry and mean field interaction 
obtained using a phenomenological $\pi\pi$ forward scattering amplitude in the elastic channels. 
The p-wave scattering with $\rho$ meson resonance 
leads to an attractive mean field interaction, 
and the presence of the absorptive part is mainly attributed to 
the formation of this resonance. 
We also incorporate a simple time dependence of the potential 
reflecting the dynamics of the evaporating source. 
Using the obtained potential, we examine how and to what extent 
the so-called HBT Gaussian radius is varied by the modification of the propagation. 
}

\begin{document}

\maketitle

\section{Introduction}

The non-Abelian nature of quantum chromodynamics manifests in the asymptotic freedom of 
%the interactions among 
quarks and gluons at high energy or short distance. 
While the colored quanta are, in ordinary circumstances, confined in color-neutral hadrons, 
they are liberated in extreme environments at high energy density. 
The study by the ultrarelativistic heavy ion collision enables us to explore 
the theoretically predicted new state of matter called the quark-gluon plasma.

The hydrodynamical simulations at the RHIC energy successfully reproduce the single-particle spectra 
by taking into account the radial flow in the central collisions 
and the elliptic flow in the peripheral collisions, 
which is regarded as an indication of the formation of the locally thermalized state. 
To extract the space-time structure of the matter, 
the two-particle momentum distribution of mesons has been measured 
to apply the technique called Hanbury Brown and Twiss (HBT) interferometry. 
This method was originally developed in microwave astronomy 
to measure the stellar radii\cite{HBT56}. 
The interference of a photon pair found in the observation was stimulating 
enough to play a seminal role in developing quantum optics\cite{Glauber63}. 
A similar interference of an identical particle pair was found in the product of 
$p\bar p$ collisions\cite{Gol60}, 
and has been utilized to measure the space-time extension of the hadron source 
created in the collider experiments\cite{YK78,GKW79}.

In the ultrarelativistic heavy ion collision, 
the space-time structure has been measured exclusively using the two-particle spectrum of identical pions. 
To extract the dynamics of the source, 
an elaborate choice of the coordinate system has been employed\cite{Pra86,Ber}, 
which refers to the motion of a pair. 
One of the axes is oriented in the direction of the incident beams. 
The other two in the transverse plane are spanned by the direction of the averaged transverse momentum of the pair 
and the direction perpendicular to it. 
They are called the longitudinal, outward and sideward directions, respectively. 
%What is important for studying the dynamics using HBT interferometry consists in the fact 
An advantage of using this coordinate system is %consists in the fact 
that the outward radius reflects the temporal structure of the source as well as the spatial one, 
owing
 to the finite interval of the emission time of a pair. 
The sideward radius is separated from the temporal structure, 
and measures a bare spatial extension. 
%The above choice of the coordinate system %enables the HBT radii to 
%maximally reflects the apparent asymmetry of the transverse radii. %due to the temporal structure. 
%The above choice of the coordinate system enables the HBT radii to maximally reflect 
%the apparent asymmetry on the transverse plane, 
%as we illustrate in Appendix \ref{sec:duration}. 
%Owing to the exclusively long lifetime of the matter in the formation of the quark-gluon plasma, 
The observation of a pronounced difference in the two radii on the transverse plane is expected 
to provide a hadronic signature for the realization of the new state of matter
\cite{Pra86,BGT88,Ber,BB89}. 
Once the quark-gluon plasma is formed, 
one expects that the time evolution of the system becomes slower 
in the phase transition (or possibly, a crossover) regime, 
which reflects in the long lifetime of the hadron source, and possibly the prolonged emission time.

In prior to the RHIC experiments, 
the dynamics of the collective expansion was simulated with hydrodynamical models assuming 
the formation of the quark-gluon plasma\cite{RG9596}. % in the collisions at $\sqrt{s_{NN}}=200$ GeV\cite{RG9596}. 
The results indicate a prolonged lifetime of the matter, 
and a considerable difference in the radii, $R\out/R\side \simeq 2$. 
More elaborate models also showed the large ratios, $R\out/R\side>1$, 
respectively\cite{SBD,HT02,KH03}. 
While these models precisely reproduce many classes of observables, especially the elliptic flow, 
the measured radii unexpectedly show a systematic deviation from the theoretical predictions\cite{STAR1,PHENIX,LPSW}, 
which has been called the ``RHIC HBT puzzle". 
This work is motivated by this longstanding issue, 
and we investigate a possible origin of the deviation.\footnote{On the experimental side, a technique called the imaging method has been developed 
to avoid assuming the Gaussian parametrization of the source profile\cite{imaging,BD97}. }
%We need more systematic studies in order to investigate the origin of the deviation

Recently, we have examined the effects of a final state interaction, 
which has not been taken into account in the hydrodynamical simulations\cite{HM}. 
It has been commonly assumed in the hydrodynamical modeling that 
the frequent interactions among particles, which maintain thermal equilibrium of the system, 
become ineffective abruptly on a certain hypersurface, 
and that the emitted particles propagate freely toward the detector after that moment. 
However, a more realistic picture is that the multiple interactions take place in the vicinity of the hypersurface 
while an escaping particle traverses the cloud formed by other evaporating particles. 
In an attempt to describe these interactions in the freeze-out stage, 
we have introduced a mean field interaction into the framework of HBT interferometry\cite{Chu,HM}, 
assuming that the other conventional approximations are still valid. 
Our model contains an imaginary part of the mean field potential 
in analogy to the optical model used in nuclear reactions, 
which describes the absorption %due to the incoherent scatterings 
in the cloud. 
Focusing on the sideward radius underestimated by the hydrodynamical simulations, 
we have examined, within a schematic model of static potential, 
how the modification of the propagation after the emission reflects in 
the HBT images extracted from the asymptotic form of the two-body momentum distribution. 
We found that the images are distorted depending on the sign of the real part 
of the mean field potential: 
an attractive interaction stretches the sideward extension, while a repulsive one shrinks it. %, in contrast. 
The absorptive part of the mean field potential acts to modify the images 
in a similar way to the effect of attractive interaction 
by cutting off the particles emitted in the deep interior and backside region of the source. 
A detailed account of these results is given in \S \ref{sec:form} and \S\ref{sec:prof}. 
We note that Cramer et al. and Pratt also examined independently but using the different frameworks 
the possible distortion of the images due to the mean field interaction\cite{CMWY,Pra06}.

These results, obtained in the previous work, imply an improvement of the HBT puzzle 
by taking into account the cooperative effects of the attractive interaction and the absorption. 
They motivate us to pursue this possibility more seriously. 
A suitable framework to simulate the effects of the mean field interaction may be 
the microscopic transport approach, in which the effects are incorporated in the Vlasov term\cite{MM}. 
It will provide a consistent description of the dynamics in the hadron phase with the mean field interaction until the last collision, 
and describe its effects caused throughout the dynamics. 
Owing to the microscopic approach\cite{BD00}, 
we obtain the space-time distribution of the last collision point without parameter fitting 
for the sake of the freeze-out criterion, to which HBT interferometry is possibly sensitive. 
%As far as we know, 
To the best of our knowledge, 
the effect of the Vlasov term in the numerical simulation has not been examined.

To illustrate the effects of the mean field interaction more schematically, 
we however examine its effects in the very last stage of the freeze-out process, assuming the hydrodynamical picture. 
We estimate the magnitude of the effects focusing on the mean field interaction after the sudden thermal ``freeze-out".\footnote{
The ``freeze-out" indicates the sudden freeze-out involved in the hydrodynamical modeling, 
%In this work, 
and we study the effects of the mean field interaction after the ``freeze-out" 
which is neglected in the conventional models: 
therefore, it causes the momentum shift of the spectrum after the ``freeze-out". 
In the following, 
the terminologies, freeze-out hypersurface and freeze-out time, are used in the same sense. }
Although the system is rather dilute after the ``freeze-out", 
the effect of the mean field interaction possibly remains in the vicinity of the freeze-out hypersurface, 
since the extension of the freeze-out hypersurface obtained from the hydrodynamical simulation is much smaller than 
that of the distribution of the last collision point observed in the microscopic approach\cite{BD00}. 
Our attempt would imply more efficient effects in the denser system before the ``freeze-out". 
%if considerable effect is found in our schematic picture. 
Our results suggest that the effect of the mean field interaction should be incorporated and examined in consistent description of the dynamics, 
e.g., the transport approach mentioned above. %, the equation of state and description of the freeze-out stage in the hydrodynamic approach. 
In our current picture, HBT interferometry extracts the distorted profile of the freeze-out hypersurface 
due to the modification of the propagation after the emission.

In \S\ref{subsec:meanfield}, 
we construct a phenomenological mean field potential to improve the previous work 
and to embark on an application to the realistic dynamics in the ultrarelativistic heavy ion collision. 
To study the modification of the Gaussian radius parameters, 
we should address the time dependence of the mean field potential 
and clarify whether it is attractive or repulsive. 
Because the mean field potential diminishes in the expansion of the matter, 
we examine a time-dependent potential within simple space-time geometry of the evaporating source, 
which enables us to avoid overestimating its effect on the observables. 
The sign of the real part of the potential is determined from a microscopic view, 
assuming that the two-body scattering of the pion, which has the most abundant population, 
dominates the origin of the mean field interaction. 
We utilize the phenomenological analysis of the elastic forward $\pi\pi$ scattering amplitude\cite{PI}. 
The magnitude of the mean field interaction and the sign of the real part are not obvious so far. 
They have not been studied in detail incorporating the momentum dependence and isospin channel dependence of the elementary scattering. 
%It also depends on the phase space distribution of the medium pions. 
A relevant feature of the low-energy $\pi\pi$ scattering in hundreds of MeV 
is the formation of $\rho$ meson resonance, 
which reflects in both the real and imaginary parts of the mean field potential. 
Owing to the attractive regime below the mass of $\rho$ meson resonance, 
we actually obtain an attractive mean field interaction for the pion carrying momentum in some hundreds of MeV. 
The effect of this attractive interaction is in favor of an improvement of the HBT puzzle, as shown in preceding works\cite{CMWY,Pra06,HM}. 
%As well as the forward $\pi\pi$ scattering amplitude, 
%this result depends on the momentum distribution at the freeze-out stage: 
%and the collective motion of the expanding source 
%which is reflected through the typical center-of-mass energy of $\pi\pi$ scatterings. 

Using the obtained phenomenological potential, 
we examine the modification of the Gaussian radius parameters in \S \ref{subsec:Rhbt}. 
We find that the deviation of the sideward radius and its ratio to the outward radius 
are improved by the attractive mean field interaction in the low-momentum regime. 
The magnitude of the effect is, however, not strong enough 
to resolve the problem by the sole effect of the mean field interaction after the ``freeze-out", 
since the density of the pion in the exterior of the freeze-out hypersurface is 
not sufficiently large, owing to the dynamics of the evaporating source. 
Nevertheless, the magnitude of the effect % by this attractive interaction 
is on the same order of the individual effect obtained by some upgrades of the hydrodynamical modeling\cite{Pra09}. 
Our result suggests that the consistent description of the dynamics should be examined incorporating the mean field interaction. 
Section \ref{sec:CR} is devoted to the concluding remarks with some brief comments on the related works.

\section{Distortion of the HBT images by the mean field interaction} \label{sec:form}

In this section, we show how the modification of the propagation by the one-body mean field potential 
is incorporated in the framework of HBT interferometry. 
Along the detailed account of the preceding work\cite{HM}, 
we extend it in the two important points. 
First, we consider the time evolution of the source, 
%or more specifically, the time evolution of the freeze-out hypersurface, 
whereas the previous work is limited to the static case. 
This extension is required to examine the time-dependent mean field potential, 
and the effect of the temporal structure reflected in the difference between the outward and sideward extensions. 
Second, we incorporate the relativity in evaluating the modification of the propagation, 
since the pion created in the ultrarelativistic heavy ion collision 
carries high momentum in the relativistic regime.

\subsection{The correlation function by the density matrix formalism}

The correlation function $C$ in HBT interferometry is defined by 
\begin{eqnarray}
C(\bm{k}_1,\bm{k}_2)  = \frac{ P_2(\bm{k}_1,\bm{k}_2) }{ P_1(\bm{k_1}) P_1(\bm{k_2}) } \ ,
\label{eq:C} 
\end{eqnarray}
where $\bm{k}_1$ and $\bm{k}_2$ are the momenta of a particle pair. 
$P_1(\bm{k})$ is the probability of detecting a particle 
and $P_2(\bm{k}_1,\bm{k}_2)$ is the probability of the simultaneous detection of a pair. 
If we observe an independent pair, 
the joint probability is factorized into the form, $P_2(\bm{k}_1,\bm{k}_2)=P_1(\bm{k}_1)P_1(\bm{k}_2)$, 
and the correlation function is given by $C=1$, independent of momenta. 
The effect of the interference of a pair is found in a deviation of the correlation function from unity. 
Even in the absence of the interactions, 
we find a correlation between an identical boson (fermion) pair due to the Bose-Einstein (Fermi-Dirac) statistics\cite{Gol60}. 
The correlation function of an identical boson pair 
deviates in the upward from unity owing to the symmetrization 
with respect to the labels of particles, as we show in the following. 
In contrast, for an identical fermion pair, the correlation function deviates 
in the downward due to the minus sign appearing in the antisymmetrization. 
HBT interferometry measures the space-time structure of the source utilizing such a correlation 
originated in the statistics in quantum mechanics. 
However, the abundant particle production in the ultrarelativistic heavy ion collision 
causes the final state interactions inherent in the particle species. 
They distort the pure statistical correlations and the extracted HBT images. 
In this work, we focus on the correlation of pion pair 
and a final state interaction described on the basis of a mean field interaction. 
A one-body potential describes the interaction between a pion and the rest of the evaporating pions.

First, we recall how the correlation function $C$ is related to the source distribution function in the free case. 
We begin with the definition of the probabilities given by the statistical averages 
at the large time $t_a\rightarrow \infty$, 
\begin{eqnarray}
P_1(\bm{k}) & := & \lim_{t_a\rightarrow \infty} \langle a^{\dagger}_{\bm k }(t_a) a_{\bm k}(t_a) \rangle \label{eq:P10} \ , \\
P_2(\bm{k}_1,\bm{k}_2) & := & \lim_{t_a\rightarrow \infty}
\langle a^{\dagger}_{\bm k_1 }(t_a) a^{\dagger}_{\bm k_2}(t_a) a_{\bm k_2 }(t_a) a_{\bm k_1}(t_a) \rangle \ , \label{eq:P20}
\end{eqnarray}
where the equal time commutation relation is imposed on the creation and annihilation operator, 
\begin{eqnarray}
\left[ a_{\bm p}(t), a^{\dagger}_{\bm p^\prime}(t) \right] = \delta(\bm p -\bm p^\prime) \ . \label{eq:com}
\end{eqnarray}
The statistical average of an operator $\mathcal O $ is described using the density matrix $\hat \varrho$, 
taking the trace with any complete set of bases: 
\begin{eqnarray}
\langle \mathcal O (t) \rangle = {\rm Tr } \left[\ \hat \varrho \mathcal O (t) \ \right] \ .
\end{eqnarray}
In the definitions of the probabilities in Eqs. (\ref{eq:P10}) and (\ref{eq:P20}), 
we take the trace using the complete multiparticle state in momentum space, 
\begin{eqnarray}
1 =  \sum_{n=0}^{\infty} \frac{1}{n!} \integ d\bp _1 \cdots \integ d\bp _n \ |\bp_1 \cdots \bp_n \rangle \langle \bp_1 \cdots \bp_n | 
\end{eqnarray}
where the first term $(n=0)$ is the vacuum term, $|0 \rangle \langle 0 |$. 
The probabilities are then given by the matrix element of the reduced one-particle density matrix $\hat{\rho}(t)$ 
and that of the reduced two-particle density matrix $\hat{\rho}_2(t)$, respectively. 
We obtain them by taking the partial traces of the density matrix $\hat\varrho$, %of the whole system, 
\begin{eqnarray}
P_1(\bm{k})  & = & \lim_{t_a\rightarrow \infty}
\sum_{n=0}^{\infty} \frac{1}{n!} \integ d\bp _1 \cdots \integ d\bp _n 
\langle \bp_1 \cdots \bp_n \bk | \hat \varrho(t_a) | \bp_1 \cdots \bp_n \bk \rangle \nonumber \\
& := & \lim_{t_a\rightarrow \infty} \ \langle\bm{k} | \hat \rho(t_a) | \bm{k}\rangle  \label{eq:P1} \\
P_2(\bm{k}_1,\bm{k}_2) & = & \lim_{t_a\rightarrow \infty}
\sum_{n=0}^{\infty} \frac{1}{n!} \integ d\bp _1 \cdots \integ d\bp _n 
\langle \bp_1 \cdots \bp_n \bm{k_1} \bm{k_2} | \hat \varrho(t_a) | \bp_1 \cdots \bp_n \bm{k_1} \bm{k_2} \rangle 
\nonumber \\
& := & \lim_{t_a\rightarrow \infty} \  
\langle \bm{k_1} \bm{k_2} | \hat \rho_2(t_a) | \bm{k_1} \bm{k_2} \rangle
 \label{eq:P2}
\end{eqnarray}
where we have switched to the Schr\"odinger picture. 
Owing to the commutation relation (\ref{eq:com}) and the definition of $P_2(\bm{k}_1,\bm{k}_2)$, 
we find the symmetry for the exchange of momenta, 
\begin{eqnarray}
\langle \bm{k_1} \bm{k_2} | \hat \rho_2(t_a) | \bm{k_1} \bm{k_2} \rangle = 
\langle \bm{k_2} \bm{k_1} | \hat \rho_2(t_a) | \bm{k_1} \bm{k_2} \rangle =
\langle \bm{k_1} \bm{k_2} | \hat \rho_2(t_a) | \bm{k_2} \bm{k_1} \rangle \ , \label{eq:sym1}
%= \langle \bm{k_2} \bm{k_1} | \hat \rho_2(T) | \bm{k_2} \bm{k_1} \rangle
\end{eqnarray}
and also for the two-time exchanges, 
\begin{eqnarray}
\langle \bm{k_1} \bm{k_2} | \hat \rho_2(t_a) | \bm{k_1} \bm{k_2} \rangle = 
\langle \bm{k_2} \bm{k_1} | \hat \rho_2(t_a) | \bm{k_2} \bm{k_1} \rangle \ . \label{eq:sym2}
%= \langle \bm{k_2} \bm{k_1} | \hat \rho_2(T) | \bm{k_2} \bm{k_1} \rangle
\end{eqnarray}
In the asymptotic region where the interactions are absent, 
the two-particle density matrix element in Eq. (\ref{eq:P2}) is factorized into 
the products of the one-particle density matrix elements in Eq. (\ref{eq:P1}), 
but maintaining the relations in Eqs. (\ref{eq:sym1}) and (\ref{eq:sym2}), as 
\begin{eqnarray}
P_2(\bm{k}_1,\bm{k}_2) &=& \lim_{t_a\rightarrow \infty} \left[
 \langle\bm{k}_1 | \hat \rho(t_a) | \bm{k}_1 \rangle  \langle\bm{k}_2 | \hat \rho(t_a) | \bm{k}_2 \rangle 
 +  \langle\bm{k}_1 | \hat \rho(t_a) | \bm{k}_2 \rangle  \langle\bm{k}_2 | \hat \rho(t_a) | \bm{k}_1 \rangle 
\right] 
\nonumber\\ 
%\mathscr{B} \huge{[} \ \langle\bm{k}_i | \hat \rho(T) | \bm{k}_j \rangle \ \huge{ ]} \\
&=& P_1(\bm{k}_1) P_1(\bm{k}_2) + 
\lim_{t_a\rightarrow \infty} \left| \langle\bm{k}_1 | \hat \rho(t_a) | \bm{k}_2 \rangle \right|^2
\ . 
\label{eq:PP2}
\end{eqnarray}
In this form, we find the correlation of the pair represented by the off-diagonal matrix element. 
The interference term is semipositive definite.

We briefly comment on the case of an identical fermion pair. 
Instead of the commutation relation (\ref{eq:com}), we impose the anticommutation relation to 
the creation and the annihilation operator, 
\begin{eqnarray}
\left\{ a_{\bm p}(t), a^{\dagger}_{\bm p^\prime}(t) \right\} = \delta(\bm p -\bm p^\prime) \ . \label{eq:Acom}
\end{eqnarray} 
We then find the relations similar to Eq. (\ref{eq:sym1}), but with opposite signs: 
\begin{eqnarray}
\langle \bm{k_1} \bm{k_2} | \hat \rho_2(t_a) | \bm{k_1} \bm{k_2} \rangle = 
- \langle \bm{k_2} \bm{k_1} | \hat \rho_2(t_a) | \bm{k_1} \bm{k_2} \rangle =
- \langle \bm{k_1} \bm{k_2} | \hat \rho_2(t_a) | \bm{k_2} \bm{k_1} \rangle \ . \label{eq:Asym1}
%= \langle \bm{k_2} \bm{k_1} | \hat \rho_2(T) | \bm{k_2} \bm{k_1} \rangle
\end{eqnarray}
On the other hand, exchanging the momenta twice, 
we still find the same relation as in Eq. (\ref{eq:sym2}) without the sign flip, 
\begin{eqnarray}
\langle \bm{k_1} \bm{k_2} | \hat \rho_2(t_a) | \bm{k_1} \bm{k_2} \rangle = 
\langle \bm{k_2} \bm{k_1} | \hat \rho_2(t_a) | \bm{k_2} \bm{k_1} \rangle \ . \label{eq:Asym2}
%= \langle \bm{k_2} \bm{k_1} | \hat \rho_2(T) | \bm{k_2} \bm{k_1} \rangle
\end{eqnarray}
If we factorize the two-particle density matrix element in Eq. (\ref{eq:P2}) 
with the relations in Eqs. (\ref{eq:Asym1}) and (\ref{eq:Asym2}) being preserved, 
we obtain the interference term for an identical fermion pair, 
\begin{eqnarray}
P_2(\bm{k}_1,\bm{k}_2) &=& \lim_{t_a\rightarrow \infty} \left[
 \langle\bm{k}_1 | \hat \rho(t_a) | \bm{k}_1 \rangle  \langle\bm{k}_2 | \hat \rho(t_a) | \bm{k}_2 \rangle 
 -  \langle\bm{k}_1 | \hat \rho(t_a) | \bm{k}_2 \rangle  \langle\bm{k}_2 | \hat \rho(t_a) | \bm{k}_1 \rangle 
\right] 
\nonumber\\
%\mathscr{B} \huge{[} \ \langle\bm{k}_i | \hat \rho(T) | \bm{k}_j \rangle \ \huge{ ]} \\
&=& P_1(\bm{k}_1) P_1(\bm{k}_2) - 
\lim_{t_a\rightarrow \infty} \left| \langle\bm{k}_1 | \hat \rho(t_a) | \bm{k}_2 \rangle \right|^2
\ .
\end{eqnarray}
The negative sign in the interference term reflects the statistics of the fermion pair.

Considering pion pair, 
the objects we have to examine are the matrix elements of the one-particle reduced matrix 
in Eqs.(\ref{eq:P1}) and (\ref{eq:PP2}). 
Keeping the Schr\"odinger picture, 
the effects of the interaction after the decoupling from the source are 
included in the time evolution of the reduced one-particle density matrix. 
By using the time evolution operator $\hat U(t_a,t_0)$, the density matrix is represented by 
\begin{eqnarray}
\hat \rho(t_a) = \hat U(t_a,t_0) \hat \rho(t_0) \hat U^\dagger (t_a,t_0) \ , \label{eq:rho}
\end{eqnarray}
where $t_0$ is the decoupling time when a particle is emitted from the source, 
and is regarded as the initial time of the propagation toward the asymptotic region. 
%The effects of the mean field interaction is incorporated in the time evolution. 
If we assume the free propagation after the decoupling, 
Eq. (\ref{eq:rho}) is explicitly written as 
\begin{eqnarray}
\hat \rho(t_a) = e^{ - i \hat H_0 (t_a-t_0) } \hat \rho(t_0) e^{ i \hat H_0 (t_a-t_0) } \ ,
\end{eqnarray}
where $\hat H_0$ is the one-particle free Hamiltonian.

Inserting the relation (\ref{eq:rho}) 
and the complete coordinate bases into Eqs. (\ref{eq:P1}) and (\ref{eq:PP2}), 
we have 
\begin{eqnarray}
\hspace{-1cm}
P_1(\bk) & =  &
\lim_{t_a\rightarrow \infty} \ \integ d\bx_1 \!\! \integ d\bx_2 \langle\bm{k} |\hat U(t_a,t_0)|\bx_1 \rangle 
\langle \bx_1 | \hat \rho(t_0) | \bx_2 \rangle \langle \bx_2 | \hat U^\dagger (t_a,t_0) | \bm{k}\rangle  
\ ,
\label{eq:Q1} \\
\hspace{-1cm}
P_2(\bm{k}_1,\bm{k}_2) &=& 
P_1(\bm{k}_1) P_1(\bm{k}_2) 
 + 
\lim_{t_a\rightarrow \infty} 
\left| 
 \integ d\bx_1 \!\! \integ d\bx_2 \langle\bm{k}_1 |\hat U(t_a,t_0)|\bx_1 \rangle \right.\nonumber\\
 && \hspace{4.5cm} \times \left.
\langle \bx_1 | \hat \rho(t_0) | \bx_2 \rangle \langle \bx_2 | \hat U^\dagger (t_a,t_0) | \bm{k}_2 \rangle 
 \right|^2
 \ .
 \label{eq:Q2}
\end{eqnarray}
The above expressions are described using two pieces: 
the density matrix element at the decoupling time, $\rho(t_0,\bx_1,\bx_2)$, 
and the amplitude of the propagation initiating at $\bx$ and terminating with the asymptotic momentum $\bk$, 
\begin{eqnarray}
\varphi_{\bk} (t_0, \bm{x};t_a ) = \langle\bm{k} |\hat U(t_a,t_0)|\bx \rangle \ . \label{eq:def_amp}
\end{eqnarray}
%where we have abbreviated to write the time $t_a$ on the left-hand side. 
By using a set of coordinates defined by $\bx = (\bx_1 + \bx_2)/2$ and $\br = \bx_1-\bx_2$, 
the Wigner function $f(t_0; \bx, \bp)$ is given by 
\begin{eqnarray}
f(t_0, \bx, \bp) = \integ d\br e^{-i \bp \cdot \br} \rho(  t_0 ; \bx + \frac{\br}{2},  \bx - \frac{\br}{2})
\ , \label{eq:def_wigner}
\end{eqnarray}
which is the quantum analog of the phase space distribution function at $t=t_0$.

In advance to embarking on the interacting case, 
we derive the familiar form of the probabilities under the absence of final state interactions. 
Without the interaction, the amplitude in Eq. (\ref{eq:def_amp}) is simply given by a plane wave 
$$\varphi_{\bk} (t_0, \bm{x};t_a) = e^{-iE_{\bk}(t_a-t_0)} e^{-i\bk\cdot\bx} \ ,$$
where the energy $E_\bk$ is the eigenvalue of the free Hamiltonian, $\hat H_0 |\bk\rangle = E_{\bk} |\bk\rangle$. 
Substituting the above plane wave and the Wigner function defined in Eq. (\ref{eq:def_wigner}), 
we obtain the expression of $P_1(\bk)$, 
\begin{eqnarray}
P_1(\bk) = \integ d\bx \!\! \integ d\bp \ \delta(\bp-\bk) f(t_0, \bx, \bp)  = \integ d\bx f(t_0, \bx, \bk) \ ,
\label{eq:P1free0}
\end{eqnarray}
where the delta function, in the intermediate of the above equation, represents the momentum conservation 
in the free motion. 
%Eq.(\ref{eq:P1free0}) represents what we expect intuitively. 
For the static source or the instantaneous emission at $t=t_0$, 
the momentum distribution at $t_0$ is preserved in the asymptotic region, 
if the interaction is absent during the propagation. 
The decoupling time in general depends on the location of the emission point, $t_0(\bx)$, 
e.g., for our interest, owing to the expansion of the source created in the ultrarelativistic heavy ion collision. 
In this case, $P_1(\bk)$ is given by collecting the contribution of the particles emitted on the hypersurface in the space-time, 
and we then obtain an extended form 
\begin{eqnarray}
P_1(\bk) & = & \integ d^4x \ \delta( t-t_0(\bx) ) \ f(t, \bx, \bk) 
\nonumber \\
& = & \integ d^4x \ S(x, \bk) \ ,
\label{eq:P1free}
\end{eqnarray}
where, motivated by the hydrodynamical model, 
we have defined the source function on the freeze-out hypersurface, 
$S(x, \bk) := \delta( t-t_0(\bx) )  f(t, \bx, \bk) $. 
\if 0
\begin{eqnarray}
P_1(\bk) & = & \integ d^4x \  S(x, \bk)  
\nonumber \\
S(t, \bx, \bk)  &:= & \delta( t-t_0(\bx) ) f(t, \bx, \bk) 
\label{eq:P1free}
\end{eqnarray}
\fi

Performing a similar procedure for $P_2(\bm{k}_1,\bm{k}_2)$ in Eq. (\ref{eq:Q2}), 
we find a simple form of the interference term given by the Fourier transform of the source function, 
\begin{eqnarray}
P_2(\bm{k}_1,\bm{k}_2) = P_1(\bm{k}_1) P_1(\bm{k}_2) 
 + 
\left|  \integ d^4x \ S(x, \bk) \ e^{i q x } \right|^2 \ ,
\end{eqnarray}
where the momenta appearing in the above relation are the averaged momentum of the pair, 
$k^\mu = (k_1 ^\mu + k_2 ^\mu) /2  $, 
and the relative momentum, $q^\mu = k_1 ^\mu - k_2 ^\mu $. %, respectively. 
Note that we adopt a convention of the metric such that 
the inner product of four vectors is given by $AB=A^0B^0-\bm{A}\cdot\bm{B}$. 
We then obtain the familiar form of the correlation function, 
\begin{eqnarray}
C(\bk, \bq) = 1 + \eta^2(\bk, \bq)
\left|  \integ d^4x \ S(x, \bk) \ e^{i q x } \right|^2 \ ,
\label{eq:Cfree}
\end{eqnarray}
where the normalization $\eta(\bk, \bq)$ is given by 
\begin{eqnarray}
\eta^{-2}(\bk, \bq) = P_1(\bm{k}_1) P_1(\bm{k}_2) \simeq P_1^2(\bk) \ . \label{eq:norm}
\end{eqnarray} 
The approximation in the last equality is valid for the small $\bq$ regime in which the correlation arises. 
In the limit of vanishing relative momentum, %$\bq\rightarrow 0$, 
the correlation function behaves as $C( \bk, 0) = 2$ for chaotic source, 
and it has the asymptotic form $C( \bk, \infty) = 1$.

Note that the relative momentum $q^\mu$ has only three independent components, and a dependent one. 
Owing to the on-shell conditions of the asymptotic momenta, 
\begin{eqnarray}
\left\{
\begin{array}{l}
E_{\bk_1}^2 = \bk_1^2 + m^2 \ ,\\
E_{\bk_2}^2 = \bk_2^2 + m^2 \ ,
\end{array}
\right.  \nonumber
\end{eqnarray}
we have a relation among the components, 
\begin{eqnarray}
q^0 = \frac{\bk}{E_\bk} \cdot \bq = \bm v_\bk \cdot \bq \ ,
\label{eq:OnShell}
\end{eqnarray}
where the velocity $\bm v_\bk$ is defined by the averaged momentum, $\bm v_\bk = \bk / E_\bk$. 
The zeroth component of the averaged momentum is approximated to satisfy the on-shell condition, 
and is denoted by $ k^0 \simeq E_\bk = \sqrt{\bk^2+m^2}  $. 
Owing to the above constraint, 
what we obtain by measuring the correlation function in Eq. (\ref{eq:Cfree}) is thus not 
the four-dimensional image of the source, but a sort of projection of the image on the three dimensions. 
Using Eq. (\ref{eq:OnShell}), 
we conventionally eliminate the zeroth component of the relative momentum. 
The correlation function in Eq. (\ref{eq:Cfree}) is then rewritten as 
\begin{eqnarray}
C(\bk, \bq) & = & 1 + \eta^2(\bk, \bq)
\left|  \integ d^4x \ S(x, \bk) \ e^{ - i \bq \cdot ( \bx - \bm v_\bk t ) } \right|^2 
\label{eq:Cfree2} \\
& = & 
1 + \eta^2(\bk, \bq)
\left|  \integ d\bx^\prime \!\! \integ dt \ S(t, \bx^\prime + \bm v_\bk t, \bk) \ 
e^{ - i \bq \cdot \bx ^\prime } \right|^2 \ ,
\end{eqnarray}
where, on the second line, we have defined the shifted coordinate, $\bx^\prime = \bx - \bm v_\bk t$. 
In this choice, we observe the image of a ``static" source, 
\begin{eqnarray}
S(\bx^\prime, \bk) =  \integ dt \ S(t, \bx^\prime + \bm v_\bk t, \bk) \ , 
\label{eq:St}
\end{eqnarray}
where the explicit time dependence of the original source function 
and the implicit dependence in the shift of the spatial coordinate, $\bm v_\bk t$, are integrated out. 
The emission at $t$ on $\bx$ is embedded in the three-dimensional image 
as if it occurs on the shifted position, $\bx^\prime = \bx - \bm v_\bk t$. 
As long as we measure the three-dimensional image by HBT interferometry, 
these two emissions cannot be distinguished\cite{Pra86}. 
%the particle propagates the distance $\bm v_\bk t$ in the time interval, $t$. 
As we mentioned in the Introduction, 
an elaborate choice of the Cartesian coordinate, %one of the axes of which is parallel to $\bk$, 
with the outward axis being parallel to $\bk$, 
enables us to separate the effect of the finite duration of the emission. 
In this frame, the velocity is represented in components as $\bm v_\bk = (v_{\perp},0,v\lon )$, 
and the shift of the coordinate is given by 
\begin{eqnarray}
\left\{
\begin{array}{l}
x^\prime \out = x\out - v_{\perp} t  \ , \\ 
x^\prime \side = x\side  \ , \\
x^\prime \lon = x\lon - v\lon t \ , 
\end{array}
\right. 
\end{eqnarray}
where the subscripts stand for the outward, sideward and longitudinal directions, respectively. 
Roughly, the outward is parallel to the line of sight, 
and the temporal interval of the emissions reflects in the apparent depth of the measured image.

In the conventional analyses, 
the spatial Gaussian profile of the source has been assumed. 
For a static source, the source function in Eq. (\ref{eq:St}) is assumed to be 
\begin{eqnarray}
S(\bx, \bk) =  V_T \exp \left( 
- \frac{x\out^2}{2 R\out ^2} 
- \frac{x\side^2}{2 R\side ^2}
- \frac{x\lon^2}{2 R\lon^2}
\right) 
\ , \label{eq:Sgauss}
\end{eqnarray}
where $V_T = \integ dt$ is the temporal volume, which is canceled in the correlation function 
owing to the same factors in $P_2(\bk_1,\bk_2)$ and $P_1(\bk_1) P_1(\bk_2)$. 
To obtain a set of useful formulae for the variances, 
we consider the second moments of the coordinate $\bx$, 
regarding the correlation function in Eq. (\ref{eq:Cfree2}) as a generating function. 
Comparing the cases with and without the Gaussian parametrization, 
we obtain 
\begin{eqnarray}
\left\{
\begin{array}{l}
R\out^2 = %\langle \tilde{x} ^2 \rangle - 2 v_\perp \langle \tilde{x}\tilde{t} \rangle + v_\perp^2 \langle \tilde{t}^2 \rangle\\
\langle (\tilde{x}\out - v_\perp \tilde t) ^2 \rangle \ , \\
R\side^2 = \langle \tilde{x}\side ^2 \rangle \ , \\
R\lon^2 = \langle \tilde{x}\lon ^2 \rangle \ , 
\end{array}
\right. \label{eq:Rfree}
\end{eqnarray}
where, as in the conventional analysis, we have chosen the frame in which we have $v\lon=0$ as well as $v\side=0$. 
The expectation value $\langle \mathscr{O} \rangle$ 
and the deviation $\tilde{x}$ from the center of the distribution are defined by 
\begin{eqnarray} 
\tilde{x} &=& x - \langle x \rangle  \ , 
\label{eq:TILfree} \\ 
\langle \mathscr{O} \rangle & = &  \frac{ \integ \ \mathscr{O} \ S(x, \bk) \ d^4x  } {\integ \ S(x, \bk) \ d^4x  } 
\ . \label{eq:EXPfree}
\end{eqnarray}
In the last part of the next section, 
we find a modification of the formulae in Eq. (\ref{eq:Rfree}) to Eq. (\ref{eq:EXPfree}) by the mean field interaction.

\subsection{The effects of mean field interaction} \label{subsec:form2}

In the previous section, we have recalled how the Wigner function, or the density matrix, is related to 
the correlation function under the conventional approximations: 
the factorization of the two-particle reduced density matrix 
and the absence of the final state interactions. 
Note that the plane waves of the free propagations result in 
the kernel of the Fourier transform in Eq. (\ref{eq:Cfree}). 
In this section, we evaluate the one-body amplitude, 
$ \varphi_{\bm k} (t_0, \bm{x}; t_a) = \langle \bm{k} | e^{-i\hat H (t_a-t_0)} | \bm{x}\rangle $, 
of which time evolution is given by the one-body Hamiltonian with the mean field interaction, 
and examine how the HBT image extracted by the integral transform is distorted. 
We consider a nonrelativistic Hamiltonian, 
\begin{eqnarray}
\hat H = \frac{\bp^2}{2m} + V(\bx) \ ,
\label{eq:H_non}
\end{eqnarray}
with a central scalar potential $V(\bx)$ that is assumed to be generated in the central collisions. 
%We assume that the mass, $m$, in the Hamiltonian is the physical pion mass. 
While the nonrelativistic Hamiltonian is sufficient for heavy particles, 
HBT interferometry in heavy ion collision often uses the spectrum of the pion 
in some hundreds MeV, or even higher momentum regime. 
In this regime, 
the amplitude of the lightest hadron with $m\simeq 140 \ {\rm MeV}$ should be evaluated with the relativistic theory. 
To improve this point, we extend the formalism using the proper time formalism\cite{Fey50}. 
This extension in the evaluation of the amplitude is carried out in parallel to the nonrelativistic case, 
and the detailed account of this point is given in Appendix A. 
Here, we consider the nonrelativistic Hamiltonian in Eq. (\ref{eq:H_non}) for simplicity.

To examine the modification of the propagation, 
we adopt a semiclassical approach, 
applying the stationary phase approximation to the path integral form of the amplitude\cite{HM}. 
The standard prescription of the path integral is given in the form 
that represents the transition amplitude of a particle 
from an initial position $\bm x$ to a final position $\bm x_{a}$. 
By evaluating it using the saddle point method, 
the action $S_c$ of the classical trajectory provides the amplitude,  
\begin{eqnarray}
\langle \bm{x}_{a} | e^{-i\hat H (t_a-t_0)} | \bm{x}\rangle \simeq 
{\cal A}(\bm x_{a},t_a;\bm x, t_0) \ e^{i {\cal S}_c(\bm x_{a},t_a;\bm x, t_0)}  \ ,
\label{eq:xx}
\end{eqnarray}
where the classical action and trajectory are specified by the boundary conditions, 
$\bm x(t_0)$ and $\bm x_{a}(t_a)$. 
%The prefactor ${\cal A}(\bm x_{\rm f},t;\bm x,0) $ is assumed to weakly depend on the coordinates. 
%and we exclusively focus on the distortion of the phase in entire this work. 
Recall that $ \varphi_{\bm k} (t_0, \bm{x}; t_a)$ is 
the amplitude to detect a particle with the asymptotic momentum $\bm{k}$ at a sufficiently large time, $t_a$. 
This amplitude is, however, proportional to the amplitude given by Eq. (\ref{eq:xx}) 
as long as the classical trajectory is uniquely specified 
by the boundary conditions $\bm x$ at $t_0$ and $\bm k$ at $t_a\rightarrow\infty$. 
We prove this point in the following.

We insert a complete set of coordinate bases into the one-body amplitude as 
\begin{eqnarray}
\langle \bm{k} | e^{-i\hat H (t_a-t_0)} | \bm{x}\rangle = \integ d\bm{x}_{a} \ 
e^{-i\bm{k}\cdot\bm{x}_{a}} \  \langle \bm{x}_{a} | e^{-i\hat H (t_a-t_0)} | \bm{x}\rangle \;,
\label{eq:kx}
\end{eqnarray}
and adopt the semiclassical approximation in Eq. (\ref{eq:xx}) on the right-hand side of the above relation. 
The classical action, ${\cal S}_c(\bm x_{a},t_a;\bm x,t_0)$, is decomposed into two parts, 
\begin{eqnarray}
{\cal S}_c(\bm x_{a},t_a;\bm x, t_0) = 
{\cal S}_c(\bm x_{a},t_a;\bm x_{\rm m},t_{\rm m}) + {\cal S}_c(\bm x_{\rm m},t_{\rm m};\bm x, t_0) \ ,
\label{eq:SS0}
\end{eqnarray}
where $\bm x_{\rm m}(t_{\rm m})$ is a position on the classical trajectory. 
If we choose $\bm x_{\rm m}$ sufficiently far outside the range of the mean field potential, 
the first term is given by the classical action of the free motion, 
\begin{eqnarray}
{\cal S}_c(\bm x_{a},t_a;\bm x_{\rm m},t_{\rm m}) = 
\frac{ m (\bm x_{a} - \bm x_{\rm m})^2 }{ 2 (t_a - t_{\rm m}) } \;, \nonumber
\end{eqnarray}
and, substituting this form for Eq. (\ref{eq:kx}) 
and carrying out the integral with respect to $\bm x_{\rm m}$, 
we obtain an expression for the amplitude, 
\begin{eqnarray}
\langle \bm{k} | e^{-i\hat H (t_a-t_0)} | \bm{x}\rangle = {\cal A} 
e^{-i \bm{k}\cdot\bm x_{\rm m} - i E_{\bm k}(t_a-t_{\rm m}) 
+ i {\cal S}_c(\bm x_{\rm m},t_{\rm m};\bm x, t_0)} \;.
\label{eq:Sx}
\end{eqnarray}
If we assume that the amplitude, $\varphi_{\bm k} (t_0, \bm{x};t_a)$, is written in the form 
\begin{eqnarray}
\varphi_{\bm k} (t_0, \bm{x};t_a) = A_{\bm k} (t_0, \bm{x};t_a) \ e^{iS_{\bm k} (t_0, \bm{x};t_a)} 
\label{eq:phi_c}\;,
\end{eqnarray}
a comparison to Eq. (\ref{eq:Sx}) leads to a formula, 
\begin{eqnarray}
S_{\bm k} (t_0, \bm{x};t_a) &=& - \bm{k}\cdot\bm x - 
\delta {\cal S}_c(\bm x_{\rm m},t_{\rm m};\bm x, t_0)  -  E_{\bm k} ( t_a - t_0 )
\label{eq:SS}\\
\delta{\cal S}_c(\bm x_{\rm m},t_{\rm m};\bm x, t_0) &:=& {\cal S}_c(\bm x_{\rm m},t_{\rm m};\bm x, t_0)
 - \left\{ \bm{k}\cdot(\bm x_{\rm m} - \bm x) -  E_{\bm k} ( t_{\rm m} - t_0 )  \right\}
\end{eqnarray}
and $$A_{\bm k} (t_0, \bm{x};t_a) = \sqrt{2\pi i(t-t_{\rm m})/m}^2 {\cal A}(\bm x_{a},t_a;\bm x, t_0) \ ,$$
where $\delta{\cal S}_c(\bm x_{\rm m},t_{\rm m};\bm x, t_0)$ is the phase shift 
caused while a particle escapes from the range of the mean field potential. 
We note that the position $\bm x_{\rm m}(t_{\rm m})$ can be arbitrarily chosen 
on the classical trajectory, as long as $t_{\rm m}$ is taken to be large 
so that the momentum becomes converged to the asymptotic momentum $\bm k$. 
Our results do not depend on the dummy coordinate $\bx_{\rm m}$ at $t_{\rm m}$.

This procedure determines $S_{\bm k} (t_0, \bm{x};t_a)$, 
if the set of boundary conditions, $\bm x$ and $\bm k$, specifies a unique trajectory. 
In general, this uniqueness is, however, not guaranteed mathematically, 
and there are some cases wherein more than two trajectories satisfy the same set of boundary conditions. 
In other words, the trajectories focus at an end of them, $\bm x$. 
A similar difficulty is known in geometrical optics as the caustics. 
If we encounter this case, the standard semiclassical prescription breaks down, 
and a more elaborate treatment of the path integral is required to find a relevant solution of the amplitude. 
In our models examined in the following sections, 
the focusing of the trajectories does not arise in the source region, 
and the semiclassical approach works well.

In the presence of the imaginary part of the potential, ${\cal U}$, 
the prefactor may be given in the form 
\begin{eqnarray}
\lim_{t_a\rightarrow\infty} A_{\bm k} (t_0, \bm{x};t_a) 
= \lim_{t_a\rightarrow\infty} a(\bm k) \exp \left[\int_{t_0}^{t_a}\!\!\! dt \ {\cal U}(\bm x_c(t)) \right] \ .
\label{eq:atte}
\end{eqnarray}
%where we have taken the limit, $t\rightarrow \infty$. 
This integral is calculated along the classical trajectory, $\bm x_c(t)$, 
%specified by the boundary conditions, $\bm x$ and $\bm k$, 
determined in the existence of the real part of the potential, ${\cal V}$. 
%determined by the real part of the potential ${\cal V}$ with given $\bm x$ and $\bm k$. 
A negative imaginary part provides the attenuation of the flux of particle while escaping the interacting region.

Based on the amplitude given in Eq. (\ref{eq:phi_c}) to Eq. (\ref{eq:atte}), 
we first consider the modification of the one-body probability $P_1(\bm k)$ in Eq. (\ref{eq:Q1}). 
Using the set of coordinates, $\bm x = (\bm x_1 + \bm x_2)/2$ and $\bm r = \bm x_1 - \bm x_2$, 
as in the preceding section, we obtain 
\begin{eqnarray}
\varphi_{\bm k} (t_0, \bx_1,t_a)  &=&  \varphi_{\bm k} (t_0, \bm{x}+\frac{1}{2}\bm r, t_a)   
\simeq e^{i\frac{1}{2} \bm r \cdot \nabla S_{\bm k} (t_0, \bx;t_a) } \varphi_{\bm k} (t_0, \bx;t_a)
\label{eq:phi1}\\
\varphi_{\bm k} (t_0, \bm{x}_2; t_a)  &=&  \varphi_{\bm k} (t_0, \bm{x}-\frac{1}{2}\bm r; t_a)  
\simeq e^{-i\frac{1}{2} \bm r \cdot \nabla S_{\bm k} (t_0, \bx;t_a) } \varphi_{\bm k} (t_0, \bx;t_a)
\label{eq:phi2}\;,
\end{eqnarray}
where the differential operator $\nabla$ acts on the coordinate $\bm x$, 
and we have neglected the higher order derivatives of $S_{\bm k} (t_0, \bx;t_a)$. 
This procedure is allowed for the potential that has a smooth spatial profile. 
This condition is satisfied automatically if the semiclassical approach itself is valid; 
the real part of the potential, ${\cal V}$, should be a smooth function 
of the spatial coordinate over a wavelength of the particle. 
Owing to Eq. (\ref{eq:SS}), the first derivative of the classical action, $\nabla S_{\bm k} (t_0, \bx;t_a)$, 
provides the initial momentum $\bm p_0$ at $\bm x(t_0)$ as 
\begin{eqnarray}
\nabla S_{\bm k} (t_0, \bx;t_a) = \nabla {\cal S}_c(\bm x_{\rm m},t_{\rm m};\bm x, t_0) = \bp_0 (t_0, \bx, \bk) \ .
\label{eq:p0}
\end{eqnarray}
Inserting the above expressions, Eqs. (\ref{eq:phi1}) and (\ref{eq:phi2}), into Eq. (\ref{eq:Q1}), 
we obtain the one-body probability given by\cite{HM} 
\begin{eqnarray}
P_1(\bm k) &=& \integ d\bm x \!\!  \integ d\bm p \ 
\delta \left( \bm p - \bp_0   \right)  f(t_0, \bx, \bp)  \left| A \right|^2 \\
&=& \left| a(\bk) \right|^2 \integ d^4x  \ S(x, \bp_0  ) \ e^{-2\gamma(t,\bx,\bk)}
\;, \label{eq:P1int}
\end{eqnarray}
where $S(x,\bp)$ is the source function defined in Eq. (\ref{eq:P1free}), 
and $\gamma(t_0,\bx,\bk) = - \int_{t_0}^{\infty}\!\!\! dt \ {\cal U}(\bm x_c(t))$ 
provides the amount of the absorption. 
For the sake of simplicity, 
we have suppressed the arguments of the initial momentum $\bp_0 (t_0,\bx,\bk)$ given in Eq. (\ref{eq:p0}).

This result agrees with our expectation: 
the real part of the potential causes a shift of the momentum, 
and the imaginary part provides an attenuation factor of a particle traversing the cloud. 
Note that the shift of the momentum distribution is a classical effect 
that is also obtained from the classical treatment\cite{BBM96,Baym}. 
In the quantum description, the phase shift of the amplitude caused in the interacting region 
contains the information on the momentum shift. 
The absorption originates in the deviation from the classical trajectory 
due to the collision in the cloud. 
This effect is mainly attributed to the formation of $\rho$ meson resonance in the pion cloud, 
and may be partly due to the formation of $\Delta$ and $K^\ast$ with much rarer nucleon and kaon, respectively. 
In HBT interferometry, 
the modification of the one-body probabilities reflects in 
the normalization of the correlation function, $C$.

We proceed to the joint probability $P_2 (\bk_1,\bk_2)$, 
and examine how the interference pattern is distorted 
owing to the modification of the amplitude. 
We begin with the expression given in Eq. (\ref{eq:Q2}), 
which is obtained by the factorization of the amplitude and density matrix of the particle pair 
preserving the boson symmetry. 
In this relation, we have the interference term, 
\begin{eqnarray}
%P_2 (\bk_1,\bk_2) &=& P_1(\bk_1)P_1(\bk_2) + \left|F(\bk_1,\bk_2)\right|^2
%\nonumber\\
%F(\bk_1,\bk_2) &= &
F(\bk_1,\bk_2; t_a)= 
 \integ d\bx_1 \!\! \integ d\bx_2 \langle\bm{k}_1 |\hat U(t_a,t_0)|\bx_1 \rangle 
\langle \bx_1 | \hat \rho(t_0) | \bx_2 \rangle \langle \bx_2 | \hat U^\dagger (t_a,t_0) | \bm{k}_2 \rangle %\ ,
  \label{eq:F} 
\end{eqnarray}
in which the effects of the interaction are incorporated in the amplitudes examined above. 
Inserting the amplitudes given in Eqs. (\ref{eq:phi1}) and (\ref{eq:phi2}) into Eq. (\ref{eq:F}), 
we find that the interference term is expressed with the difference in the classical actions 
\begin{eqnarray}
F(\bm k_1, \bm k_2;t_a) &=& a(\bm k_1)a(\bm k_2) 
\integ d\bm x \!\!  \integ d\bm p 
\delta \left( \bm p - \bp_0  \right)  f(t_0, \bx, \bp)  
\nonumber\\
&& \hspace{1.5cm} \times \ 
e^{ -\gamma(t_0, \bx, \bm k_1) -\gamma(t_0, \bx, \bm k_2) } \ 
e^{i \{  S_{\bm k_1} (t_0, \bx; t_a) -  S_{\bm k_2} (t_0, \bx; t_a) \} } 
%\ ,
 \label{eq:F0}
\end{eqnarray}
where the initial momentum $\bp_0 (t_0,\bx,\bk)$ is given by 
the derivative of the classical action as in Eq. (\ref{eq:p0}). 
Here, we expand the actions with respect to the relative momentum $\bm q$, 
assuming that the interference arises in the small $\bm q$ regime 
compared with the magnitude of the averaged momentum $\bm k$. 
This approximation is valid for the interferometry in the ultrarelativistic heavy ion collision, 
in which the averaged momentum is in hundreds of MeV, 
while the relative momentum is on the order of the inverse of the source size $\sim \!10$ fm, which is in tens of MeV. 
Up to the leading order of the relative momentum, we have 
\begin{eqnarray}
S_{\bm k_1} (t_0, \bx; t_a) - S_{\bm k_2} (t_0, \bx; t_a) 
%&=& 
%S_{\bm k + \frac{1}{2} \bm q} (t; \bm x ,t_0)
%-
%S_{\bm k - \frac{1}{2} \bm q} (t; \bm x ,t_0)
%\nonumber\\
&\simeq&   \bm q \cdot \nabla_{\bm k} S_{\bm k} (t_0, \bx; t_a) \ , \nonumber
\end{eqnarray}
where $\nabla_{\bm k}$ operates on the averaged momentum $\bm k$ explicitly, 
and the energy $E_\bk$ implicitly. 
The attenuation factor $\gamma$ is also expanded as 
$\gamma(t_0, \bx, \bm k_1) + \gamma(t_0, \bx, \bm k_2) \simeq 2\gamma(t_0, \bm x, \bm k)\;.$
Using these expressions, we obtain 
\begin{eqnarray}
F(\bm k_1, \bm k_2; t_a) = a(\bm k_1)a(\bm k_2) 
\integ d\bx  \ f(\bm x,\bm p_0) \ 
e^{-2\gamma(t_0, \bm x, \bk)} \ 
e^{i \bm q \cdot \nabla_{\bm k} S_{\bm k} (t_0, \bx; t_a) } \ .
\label{eq:F2}
\end{eqnarray}
%where $\bp_0 (t_0, \bx, \bk) = \nabla  S_{\bk} (t; \bx,t_0)$ is 
%the initial momentum as we encountered in the one-body probability $P_1(\bm k)$. 
Note that the emission occurs at the time $t_0(\bx)$ depending on the position $\bx$.

To see how the source images are distorted by the mean field interaction, 
we write $$S_{\bm k} (t_0, \bx; t_a) = S_{0\hspace{0.05cm}\bk} (t_0, \bx; t_a) + \delta S_{\bm k} (t_0,\bx)\;,$$ 
where $S_{0\hspace{0.05cm}\bk} (t_0, \bx; t_a)$ 
is the classical action of the free motion given by Eq. (\ref{eq:SS}) as 
$S_{0\hspace{0.05cm}\bk} (t_0, \bx; t_a) = - \bm k \cdot \bm x - E_{\bm k} (t_a-t_0)$, 
and $\delta S_{\bk} (t_0,\bx)$ is the phase shift caused by the mean field interaction, 
$$\delta S_{\bk} (t_0,\bx)= S_{\bk} (t_0, \bx; t_a) - S_{0\hspace{0.05cm}\bk} (t_0, \bx; t_a) \ . $$ 
The difference in the actions, or the phase shifts, does not depend on the time $t_a$, 
since the difference is generated only in the interaction region localized in the vicinity of the source. 
Taking the derivative with respect to $\bm k$, 
we find a relation 
\begin{eqnarray}
\nabla_{\bm k} S_{\bk} (t_0, \bx; t_a) = - \bx - \bm v_{\bk} (t_a-t_0) + 
\nabla_{\bm k}\delta S_{\bk} (t_0 \bx) \ , \label{eq:DS}
\end{eqnarray}
where, as in the previous section, the velocity is defined by the off-shell momentum, $\bm v_\bk = \nabla_{\bk} E_{\bk}$. 
We, however, usually assume the on-shell condition, $E_{\bk}^2 \approx \bk^2+m^2$, for the small relative momentum. 
By inserting Eq. (\ref{eq:DS}) into Eq. (\ref{eq:F2}), the interference term can be expressed as 
\begin{eqnarray}
\hspace{-1cm}
F(\bm k_1, \bm k_2; t_a) &= &
a(\bm k_1)a(\bm k_2) \ e^{ - i \bq \cdot \bm v_{\bm k} t} \nonumber\\
&& \hspace{0.5cm} \times 
\integ d\bm x  \ f(\bm x,\bm p_0) \ 
e^{-2\gamma(t_0, \bm x, \bk)} \ 
e^{ -i \bq \cdot \left( \bm x - \bm v_\bk t_0 - \nabla_{\bm k}  \delta S_{\bk} (t_0,\bx) \right) } 
\ . \label{eq:F1} 
\end{eqnarray}
Note that the derivative of the action, $\nabla_{\bm k} \delta S_{\bk} (t_0,\bx)$, 
is not the variation of the action with respect to the momentum shift on a classical trajectory, 
but originates in the difference in the two actions of the adjacent trajectories, 
labeled by the asymptotic momenta, $\bm k_1$ and $\bm k_2$.

Together with the one-body probability in Eq. (\ref{eq:P1int}), 
the correlation function defined in Eq. (\ref{eq:C}) is then given by 
\begin{eqnarray}
C  =  1 + \left| 
\frac{\integ d\bm x  \ f(t_0, \bm x,\bm p_0) \ 
e^{-2\gamma(t_0, \bm x, \bk)} \ 
e^{ -i \bq \cdot \left( \bm x - \bm v_\bk t_0 - \nabla_{\bm k}  \delta S_{\bk} (t_0,\bx) \right) } }
{
\integ d\bm x \ 
 f(t_0, \bx, \bp_0) \ e^{-2\gamma(t_0,\bx,\bk)}
}
\right| ^2 \ , \label{eq:Cint1}
\end{eqnarray}
or, using the relation (\ref{eq:OnShell}), we obtain 
\begin{eqnarray}
C = 1 + \left| 
\frac{\integ d^4x  \ S(x, \bp_0 (x,\bk) ) \ 
e^{-2\gamma(x,\bk)} \ 
e^{ i q_{\mu} \left( x^{\mu} + \partial_k ^{\mu} \delta S_{\bk} (x) \right) } }
{\integ d^4x  \ S(x, \bp_0 (x,\bk) ) \ e^{-2\gamma(x,\bk)}}
\right| ^2 
\ , \label{eq:Cint2}
\end{eqnarray}
where $x$ is the space-time coordinate, $x^{\mu}=(t_0,\bx)$. 
The source function $S(x,\bp)$ is defined in Eq. (\ref{eq:P1free}). 
In these expressions, we have approximated the momenta appearing in the denominator as 
$\bm k_1 \simeq \bm k_2 \simeq \bm k$ for the small relative momentum. 
The temporal component of the derivative $\partial_k^\mu$ operates on the energy $E_\bk$, 
and the spatial components operate only on the explicit momentum dependence $\bk$, but not on the energy $E_\bk$.

In Eq. (\ref{eq:Cint1}), or Eq. (\ref{eq:Cint2}), 
we find the two modifications by the real part of the mean field potential: 
the momentum shift in the source function due to the acceleration, 
and the coordinate shift in the kernel of the transformation given by the derivative of the classical action. 
%, $x^{\prime\mu} = x^{\mu} + \partial_k ^{\mu} \delta S_{\bk} (x)$. 
As in the one-body probability, the imaginary part provides the attenuation factor. 
Incorporating these effects and assuming the Gaussian shape of the source profile as in Eq. (\ref{eq:Sgauss}), 
we still find the set of formulae in the form given by Eq. (\ref{eq:Rfree}). 
The modifications merely reflect in the definitions of the deviation $\tilde{x}^\mu$ 
and the expectation value $ \langle \mathscr{O} \rangle $, 
\begin{eqnarray} 
\tilde{x}^\mu %&=& x^{\prime\mu}  - \langle x^{\prime\mu}   \rangle 
%\nonumber\\
&=&  \left( x^\mu + \partial_k ^{\mu} \delta S_{\bk} (x) \right)
 - \langle  x^\mu + \partial_k ^{\mu} \delta S_{\bk} (x)  \rangle 
\label{eq:COORint}\\
\langle \mathscr{O} \rangle & = &  
\frac{ \integ \ \mathscr{O} \  S(x, \bp_0 ) \ e^{-2\gamma(x,\bk)} \ d^4x  } 
{\integ \ S(x, \bp_0 ) \ e^{-2\gamma(x,\bk)} \ d^4x  } 
\ , \label{eq:EXPint}
\end{eqnarray}
in which we find the three effects mentioned above: the momentum shift, the coordinate shift and the attenuation. 
The arguments of the initial momentum $\bp_0(x,\bk)$ are suppressed in Eq. (\ref{eq:EXPint}). 
Replacing Eqs. (\ref{eq:TILfree}) and (\ref{eq:EXPfree}) by the above definitions, 
we obtain the Gaussian parameters in the presence of the mean field interaction.

To examine the distortion of the HBT image more intuitively, 
we define the apparent emission point $\bm x^{\prime}$ as 
\begin{eqnarray}
\bm x^\prime = \bm x - \bm v_\bk t_0(\bx) - \nabla_{\bm k}\delta S_\bk(t_0(\bx),\bm x) \;, \label{eq:xshift}
\end{eqnarray}
where the right-hand side of the above relation appears 
in the kernel of the integral transform in Eq. (\ref{eq:Cint1}). 
Transforming the integral variable $\bm x$ to $\bx^\prime$, 
we obtain the integral for the interference term given by 
\begin{eqnarray}
F(\bk_q,\bk_2;t_a) = a(\bm k_1)a(\bm k_2) e^{-i \bq \cdot  \bm v_{\bm k} t_a}
\integ d\bm x^\prime  \ f_{\rm eff}(\bm x^\prime, \bm k) \ 
e^{i \bm q \cdot \bm x^\prime }  \label{eq:feff}
\end{eqnarray}
with the definition, 
\begin{eqnarray}
f_{\rm eff}(\bm x^\prime, \bk) = J(\bm x, \bm x^\prime; \bm k)
\ f(t_0, \bm x,\bm p_0 ) \ e^{-2\gamma(t_0, \bm x, \bk)} \;,
\label{eq:feff2}
\end{eqnarray}
where $J(\bm x, \bm x^\prime; \bm k) = \partial(\bm x,\bm k)/\partial(\bm x^\prime,\bm k)
= \left[ \partial(\bm x^\prime,\bm k)/\partial(\bm x,\bm k) \right]^{-1}$ 
is the Jacobian of the transformation in Eq. (\ref{eq:xshift}). 
Note that the original coordinate $\bm x$ appearing in the expression of $f_{\rm eff}(\bm x^\prime, \bk)$ 
%on the right-hand side of Eq.(\ref{eq:feff}) 
is understood to be a function of $\bm x^\prime$ and $\bm k$ via Eq. (\ref{eq:xshift}). 
We have obtained the static case of the above relations in the previous work\cite{HM}, 
in which the temporal term $\bm v_{\bk} t_0(\bx)$ in Eq. (\ref{eq:xshift}) is absent. 
The modification of the propagation after the emission, arising as the phase shift of the one-body amplitude, 
is transferred to a distortion of the spatial distribution by the apparent shift of the emission point defined in Eq. (\ref{eq:xshift}). 
We find that the correlation function is written in the same form as in the free case 
using the Fourier transform, 
\begin{eqnarray}
C(\bm k, \bm q) = 1 + \left| \integ d\bm x^\prime 
 \rho_{\rm eff}(\bm x^\prime,\bm k) e^{i\bq \cdot \bx^\prime} \right| ^2
\end{eqnarray}
with the effective distribution function defined by 
\begin{eqnarray}
 \rho_{\rm eff} (\bm x^\prime,\bm k) = 
\frac{ f_{\rm eff}(\bm x^\prime, \bm k) }
{
\integ d\bm x \ 
 f(t_0, \bx, \bp_0 ) \ e^{-2\gamma(t_0,\bx,\bk)}
%\integ d\bm x \ f \left( \bm x,\bm p (\bm x,\bm k) \right) 
%e^{-2\gamma(\bm x,\bm k)} 
} \;. \label{eq:reff}
\end{eqnarray} 
Owing to the nonlinear nature of the transformation in Eq. (\ref{eq:xshift}), 
the original coordinate $\bm x$ is mapped onto the apparent coordinate $\bm x^\prime$ 
with a deformation of the grids, 
and it induces the nonlinear mapping of the source distribution function 
onto the effective one defined on the apparent coordinate system. 
This nonlinear nature acts to distort the image viewed on the coordinate $\bm x^\prime$. 
The attenuation factor provides the opacity of the cloud. %, 
%and gives a weight on the emissions on the front surface of the source. 
Integrating both sides of Eq. (\ref{eq:reff}) with respect to $\bx\prime$, 
we observe that the effective distribution function $\rho_{\rm eff}(\bm x^\prime, \bk)$ is normalized to be one at each $\bk$ 
even in the presence of the absorption, whereas its effect reduces $f_{\rm eff}(\bm x^\prime, \bk)$ in Eq. (\ref{eq:feff2}). 
We find that the normalization of the effective distribution function 
is compensated by the single-particle spectrum in the denominator of Eq. (\ref{eq:reff}), 
which contains the same attenuation factor. 
%The effective distribution function represents the distribution of the emission points of those particles 
%which survive in the interaction region to escape into the asymptotic region. 
The effective distribution function represents the distribution of the emission points of the surviving particles. 
The absorptive effect does not change the magnitude of the correlation term at the vanishing relative momentum, $C(\bk,0)=2$. 
%We find that the one-body probability $P_1(\bk)$ normalizes $f_{\rm eff}(\bm x^\prime, \bk)$, 
%and that the effective distribution function $\rho_{\rm eff}(\bm x^\prime, \bk)$ provides 
%the relative distribution of the emission point in the space-time. 
In \S \ref{sec:prof}, we illustrate the distorted profiles of the effective distribution function 
using a schematic model of the potential.

We briefly comment on the computation procedure concerning the following two points. 
In Eq. (\ref{eq:xshift}), we take the derivative of the classical action with respect to $\bm k$ 
using the cylindrical coordinate in the momentum space, 
since the conventionally used coordinate refers to the direction of $\bm k$. 
The apparent shift of the emission point is represented by its components as 
\begin{eqnarray}
\left\{
\begin{array}{l}
x^\prime = x - v_\perp t_0(\bx) - \frac{\partial \delta S_\bk (\bx)}{\partial k}  \ , \\
y^\prime = y - \frac{1}{k} \frac{\partial \delta S_\bk (\bx)}{\partial \theta_{\bm k}} \ ,  \\
z^\prime = z - \frac{\partial \delta S_\bk (\bx)}{\partial z} \ , 
\end{array}
\right. \label{eq:shiftxy}
\end{eqnarray}
where $k$, $\theta_{\bm k}$ and $v_\perp$ are the magnitude and angle of momentum $\bm k$, 
and the magnitude of the velocity $\bm v_{\bk}$, respectively. 
Owing to the definition of the coordinate system in which we have $(v_x, v_y , v_z) = (v_\perp,0,0)$, 
the temporal term proportional to the velocity is preserved only in the outward coordinate $x$. 
Even in the free case without the derivative terms, 
we find the apparent extension in the outward direction due to the presence of the temporal term. 
This effect is absent for the static source and the instantaneous emission, 
which do not have the temporal structure.

To calculate the classical action of the trajectory specified by 
the emission point $\bm x$ and the asymptotic momentum $\bm k$, 
we have to find the angle of the initial momentum $\bm p(t=0)$ such that, 
it satisfies the matching condition at the large time as 
$$\bm p(\infty) = \bm k\;.$$ 
Owing to the energy conservation law, the magnitude of the initial momentum $\bm p(0)$ 
is simply related to that of the asymptotic momentum $\bm k$. 
On the other hand, the angular momentum conservation is not adequate to fix the angle of the initial momentum, 
and we obtain the angle by explicitly solving the equation of motion. 
%a trajectory obtained with an arbitrary angle of initial momentum does not necessarily 
%converge to the asymptotic momentum $\bm k$. 
Regarding the left-hand side of the above condition as a function of $\bm x$ and $\bm p(0)$, 
we solve it with respect to the angle of $\bm p(0)$. 
A good number of the trials of computing the initial value problem, 
and their interpolation, enable us to find the solution. 
Carrying out such calculations for the adjacent values of the momentum $\bk$, 
we obtain the derivative of the phase shift, $\delta \bm x = - \nabla_{\bm k}\delta S(\bm x, \bm k)$, 
and the apparent shift of the emission point defined by Eq. (\ref{eq:xshift}).

\section{Profiles of the distorted images{\rm \cite{HM}}} \label{sec:prof}

In this section, we examine the distortion of the HBT images 
based on the effective source distribution function defined in Eq. (\ref{eq:reff}) for the static source. 
Here, the temporal term in Eq. (\ref{eq:shiftxy}) is absent. 
We show how the difference in the effects of the mean field interactions, i.e., attraction and repulsion, 
appears in the distorted images. 
Nonrelativistic analysis is performed 
using a schematic model of the mean field potential given below. 
%Due to the one-body mean field interaction in the pion cloud surrounding the source, 
%the bare distribution function $f(\bm x, \bm p)$ is distorted to become the apparent one. 
%The apparent shift of emission point $\bm x^\prime$ defined by eq.(\ref{eq:shiftxy}) 
%is computed for given emission point $\bm x$ and the asymptotic momentum $\bm k$ 
%owing to the procedure remarked below eq.(\ref{eq:reff}). 
Recall that the coordinate system on the transverse plane is spanned by 
the direction of the averaged momentum of particle pair $\bm k$ 
and the direction perpendicular to it. 
They are called outward ($x$) and sideward ($y$), respectively. 
Roughly, the particle pair is detected in the positive direction of the outward.

For simplicity, we assume the factorization of the bare source distribution function as 
\begin{eqnarray}
f(\bm x, \bm p) = \rho(\bm x) f_{{\rm th}}(\bm p;T,\mu) \ ,
\end{eqnarray}
where $\rho(\bm x)$ is the spatial profile of the source and $ f_{{\rm th}}(\bm p;T,\mu)$ is 
the thermal spectrum specified by temperature $T$ and pion chemical potential $\mu$. 
We input a two-dimensional Gaussian profile on the transverse plane, 
$\rho(\bm x) = \rho_0 \ e^{- \bm x^2/2\lambda^2}$, with $\lambda = 5$ fm 
and the Bose-Einstein distribution function with $T=140$ MeV and $\mu=0$ MeV.

We have chosen a schematic pion optical potential 
assuming a two-range Gaussian shape for the real part: 
\begin{eqnarray}
{\cal V}(\bm x) = {\cal V}_1 e^{ - \frac{\bm x^2}{2\lambda_1^2} } 
+ {\cal V}_2 e^{ - \frac{(|\bm x| - a)^2}{2\lambda_2^2} }  
\ , \label{eq:pote}
\end{eqnarray}
and another Gaussian shape for the imaginary part: 
\begin{eqnarray}
{\cal U}(\bm x) = - \ {\cal U}_0 e^{ - \frac{\bm x^2}{2\lambda_0^2} } \;.
\end{eqnarray}
The shorter ranges, $\lambda_1$ and $\lambda_0$, are taken to be consistent with 
the source profile $\lambda$ given above. 
The other range $\lambda_2$ gives the thickness of the meson cloud. 
We have chosen $\lambda_1 = 5$ fm, $\lambda_2 = 5$ fm and $a = 10$ fm; ${\cal V}_1 = \pm 10$ MeV, 
${\cal V}_2 = \pm 2$ MeV and ${\cal U}_0 = 0.1$ MeV. 
We compare the effects of the repulsive interaction and the attractive one.

\begin{figure}[t]
  \begin{center}
        \includegraphics[scale=0.9]{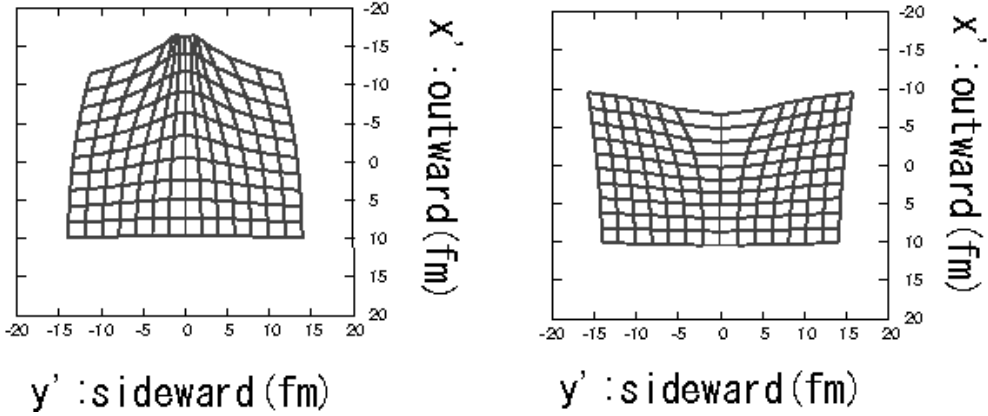}      
  \end{center}
  \caption{
Original grids mapped onto the apparent coordinate system: 
Curves on the left (right) panel indicate the deformation of the grids 
by repulsive (attractive) mean field interaction. 
We show the deformation at the asymptotic momentum $k=100$ MeV 
in the range $|\bm x| \leq 10$ fm and $|\bm y| \leq 14$ fm. 
The deformation induces the apparent distortions of the source images shown in 
Figs. \ref{fig:cntr100} and \ref{fig:cntr150}. 
}
\label{fig:grid}
\end{figure}
\begin{figure}[t]
  \begin{center}
      \includegraphics[scale=0.8]{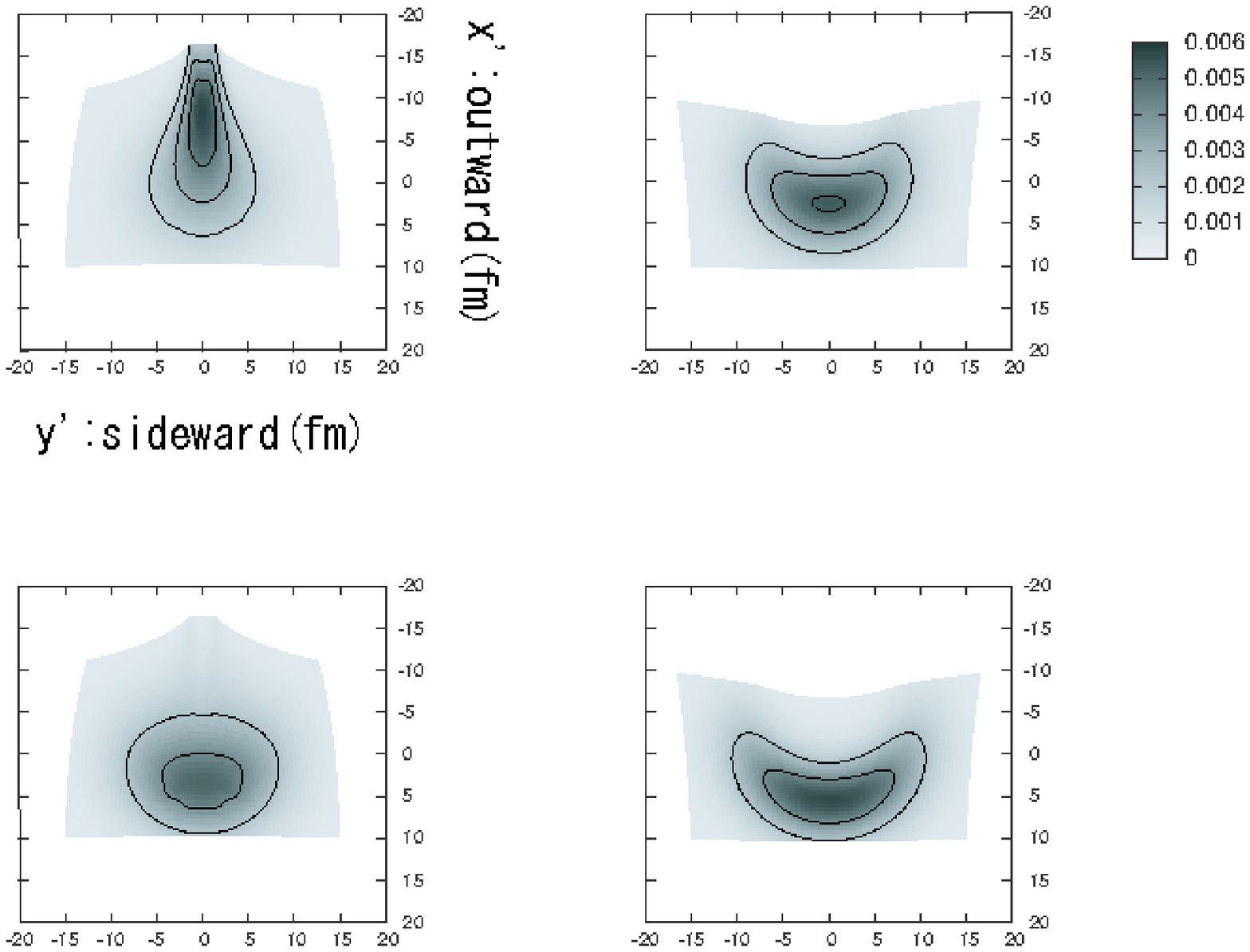}
  \end{center}
   \caption{Contour plots of the effective source distribution function, 
$\rho_{\rm eff} (\bx', \bk)$, defined in Eq. (\ref{eq:reff}).  
The averaged momentum of the pair is chosen to be $k=100$ MeV. 
The panels in the left (right) column show the distorted profiles due to 
the repulsive (attractive) interaction. We incorporate the effect of the absorption 
in the lower two panels. 
The vertical and horizontal axes are the transformed outward and sideward coordinates, respectively. 
We find the deformation of the grids by the nonlinear coordinate transform 
in the distorted perimeters of the plot range. }
   \label{fig:cntr100}
\end{figure} 
\begin{figure}[t]
  \begin{center}
        \includegraphics[scale=0.8]{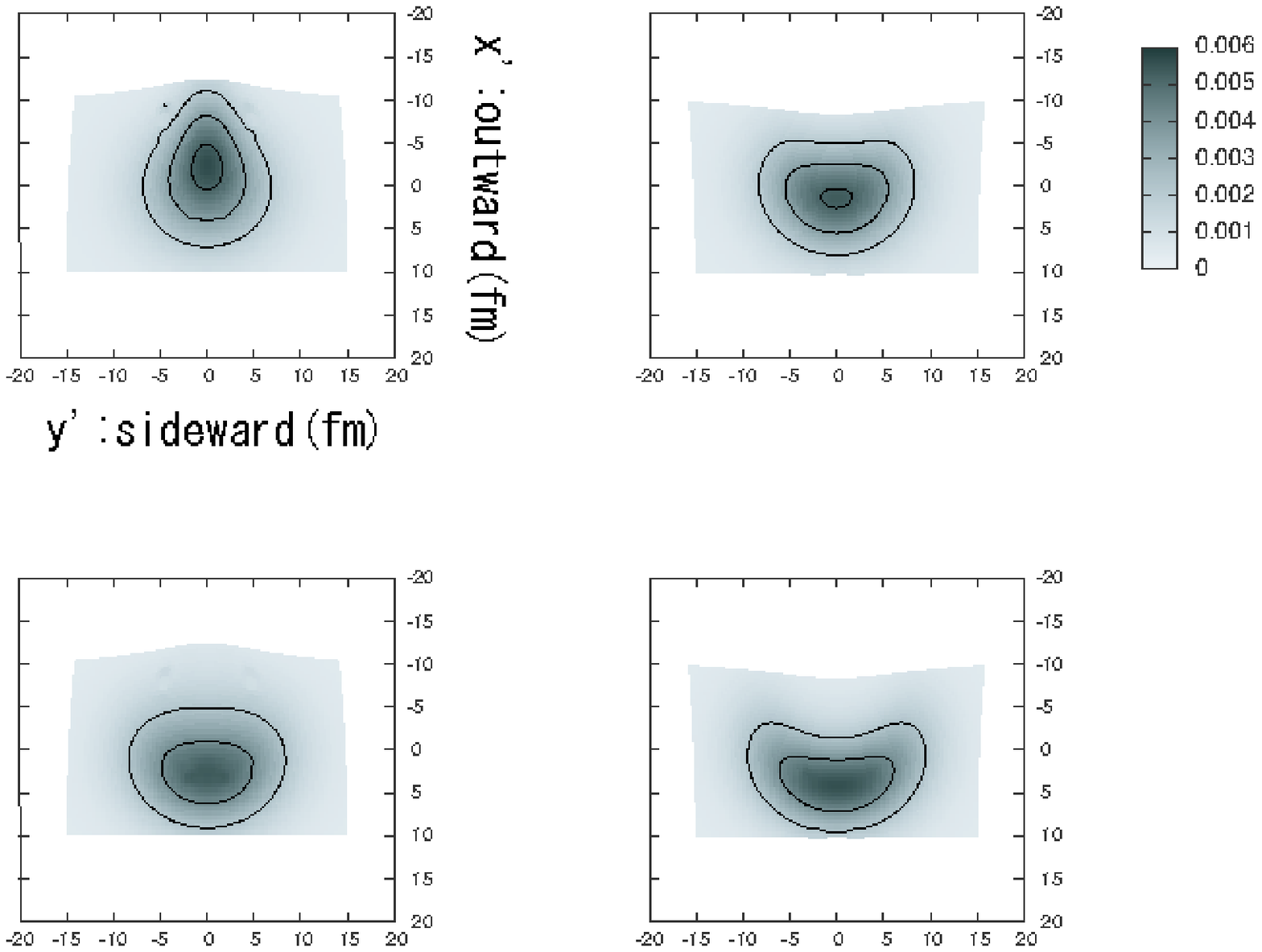}
  \end{center}
   \caption{Contour plots of the effective source distribution function, $\rho_{\rm eff} (\bx', \bk)$, at $k=150$ MeV. 
In each panel, we incorporate the same interaction as in the panel at the same location in Fig. \ref{fig:cntr100}.}
   \label{fig:cntr150}
 \end{figure}

In \S \ref{subsec:form2}, we have shown that 
the phase shift caused by the real part of the mean field potential 
results in the apparent shift of the space-time coordinate. 
We have defined a nonlinear transformation in Eq. (\ref{eq:shiftxy}), 
which maps the original coordinate system onto the apparent one. 
In Fig. \ref{fig:grid}, we show the location of the original grids 
mapped on the apparent coordinate system at the asymptotic momentum $k=100$ MeV, 
and in the range $|\bm x| \leq 10$ fm and $|\bm y| \leq 14$ fm. 
Curves on the left (right) panel represent the original Cartesian coordinate 
under the effect of the repulsive (attractive) mean field interaction. 
Induced by the deformation of the grids, 
the profiles of the image are distorted on the apparent coordinate system, 
as shown in Figs. \ref{fig:cntr100} and \ref{fig:cntr150}. 
The deformation of the grids in Fig. \ref{fig:grid} represents 
the geometrical interpretation of the effects by the real part of the potential. 
However, note that each curve does not indicate a refracted trajectory of the emitted particles, 
which is often described to express the lensing effect in geometrical optics.

Using the contour plot, 
we show the profiles of the effective distribution function defined in Eq. (\ref{eq:reff}) 
at $k = 100$ MeV in Fig. \ref{fig:cntr100}, and $k = 150$ MeV in Fig. \ref{fig:cntr150}. 
The upper panels in each figure show the distortion of images without the absorption, 
whereas the lower panels contain the effect of the absorption with ${\cal U}_0 = 0.1$ MeV 
as well as the effect of the real part of the potential. 
The two panels in the left column show the cases with the repulsive interaction, 
and those in the right column the attractive one. 
The vertical and horizontal axes extend into the apparent outward and sideward directions, respectively: 
thus, the location on each panel is the apparent emission point. 
In Figs. \ref{fig:cntr100} and \ref{fig:cntr150}, 
the distorted perimeters of the plot range indicate the deformation of the original grids.

We first compare the cases without the absorption shown in the upper row. 
At $ k =100$ MeV, 
we find that the apparent source image 
is elongated in the outward direction by the repulsive interaction, 
while it is stretched in the sideward direction by the attractive one. 
These effects are pronounced in the backside region of the source, 
since the amplitudes of the particles, penetrating into the interior of the source, are strongly distorted. 
However, the absorption damps the amplitudes of such particles 
emitted in the deep interior and backside region of the source, 
and diminishes the remarkable effects caused by the real part of the mean field potential. 
On the lower two panels, we find the images stretched in the sideward direction 
with the dominant distributions on the detector side of the source. 
The cylindrical symmetry assumed for the original source profile is broken by the interaction, 
since the strengths of these effects depend on the emission point. 
As we found in  Figs. \ref{fig:cntr100} and \ref{fig:cntr150}, 
only the reflection symmetry with respect to the outward axis survives in the presence of the mean field interaction. 
At $ k =150$ MeV, these effects are found in the same trend, but are less effective: 
the distortion of the amplitude is relatively weak at the high momentum, 
if the mean field potential does not depend on momentum.

 \begin{figure}[t]
  \begin{center}
        \includegraphics[scale=0.8]{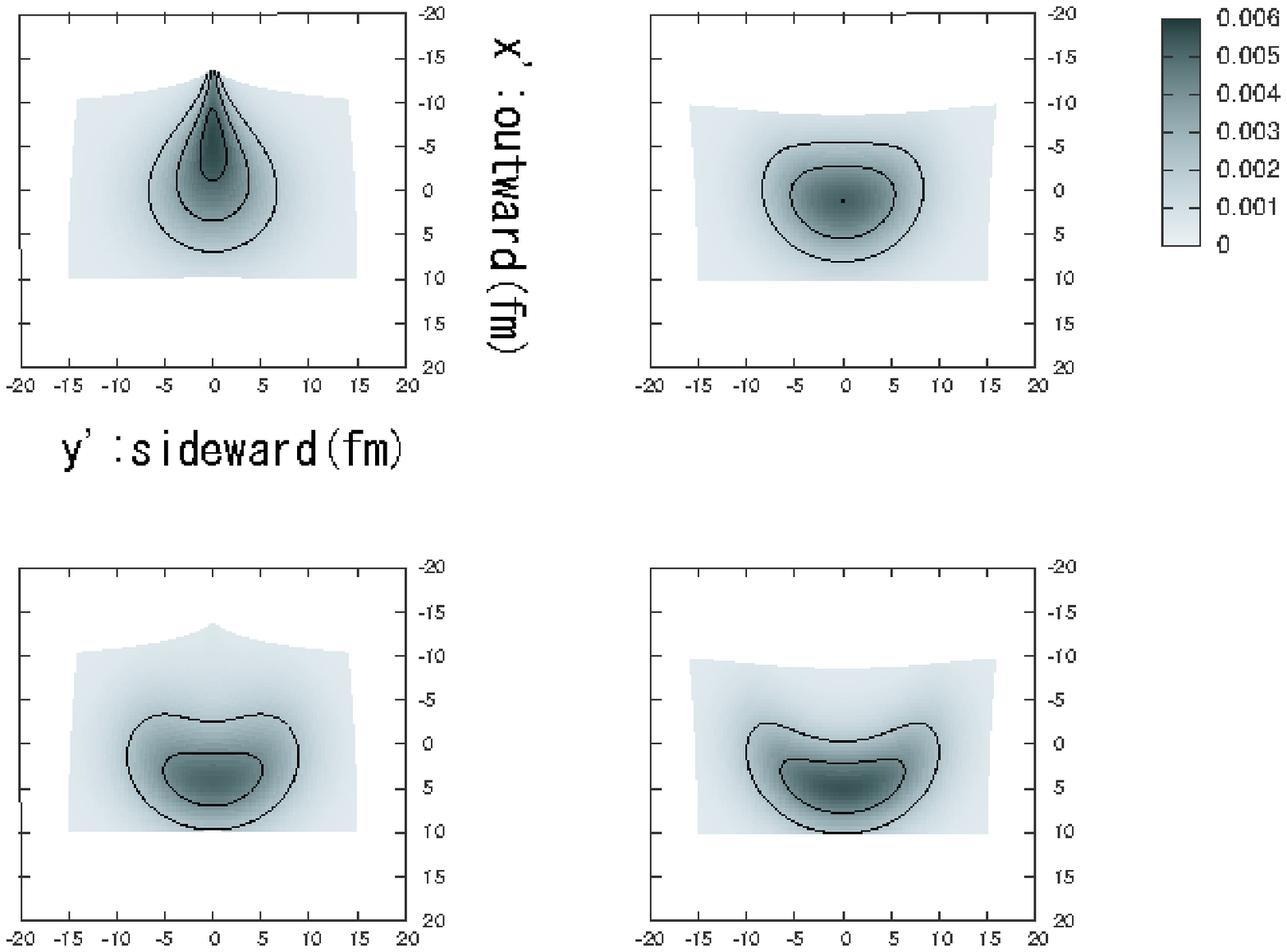}      
  \end{center}
  \caption{Effective source distribution function, $\rho_{\rm eff} (\bx', \bk)$, at $k=100$ MeV obtained with the Glauber-type approximation: 
We have chosen the same potentials as those in Fig. \ref{fig:cntr100}. 
These results qualitatively reproduce the features found in Fig. \ref{fig:cntr100}. }
   \label{fig:Glauber}
 \end{figure}

In our analysis, the origin of the distortion, by the real part of the mean field potential, is 
the nonlinear mapping of the coordinate system. 
%in Eq.(\ref{eq:shiftxy}) 
%as we have shown in Fig.\ref{fig:grid}. 
The transformation is given by the derivative of the phase shift 
with respect to the averaged asymptotic momentum, $\bm k=(\bm k_1 + \bm k_2)/2$. 
Recall that the derivative indicates 
not the variation of the action with respect to the momentum shift on a trajectory, 
but the difference in the actions of two adjacent trajectories. 
In the presence of the one-body mean field potential, 
we have different phase shifts on each trajectory labeled by $\bm k_1$ and $\bm k_2$, 
and the common emission point $\bm x$. 
Accordingly, their relative momentum is shifted by the one-body potential 
due to the coordinate dependence of the force to which each particle is subjected. 
The shift of the relative momentum is the origin of the distortion of the images observed utilizing a quantum interference. 
Thus, we remark that the distortion of the images in HBT interferometry is not 
a direct analogue of that in geometrical optics described on the basis of refracted trajectories.

To emphasize an aspect of the quantum interference, 
we examine the phase shift using a Glauber-type approximation 
without the refraction of trajectories. 
It assumes the straight line trajectory in the interaction region, 
so that the phase shift is given by a simple formula: %which may be written for $V << \bk^2/2m$: 
\begin{eqnarray}
\delta S_{\rm Glauber} (\bm x,\bk) 
&\simeq& - \frac{m}{|\bk|} \int_{u (0)}^{u (T)} V(\sqrt{u^2 + b^2}) \ du \ ,
\label{eq:SGlauber}
\end{eqnarray}
where $b = \sqrt{ \bx^2 - (\bx \cdot \bk)^2/ \bk^2}$ is the ``impact parameter" of the trajectory 
defined by the distance between the straight line and the center of the potential. 
Position, $u (t) = \bx(t) \cdot \bk/ |\bk|$, is the coordinate along the straight line at time, $t$. 
We then obtain the coordinate transformation (\ref{eq:shiftxy}) in the outward direction 
from the dependence on the magnitude of momentum, 
and in the sideward direction from the dependence on argument $\theta_\bk$ through $b$. 
The distortion of images at $k=100$ MeV is shown in Fig. \ref{fig:Glauber}, 
using the same potentials as in Fig. \ref{fig:cntr100}. 
We find the qualitatively same results assuming the straight line trajectory. 
%The effect of the mean field interaction is reflected in the images via shift of the relative momentum. 

\section{Mean field interaction in the freeze-out stage}

\subsection{Phenomenological mean field interaction} \label{subsec:meanfield}

In this section, we examine the dispersion relation of the pion in the cloud 
formed by the other evaporating particles. 
In analogy to the optical model employed in the analysis of nuclear reactions, 
we construct a phenomenological self-energy based on the two-body forward scattering amplitude. 
Although an escaping pion possibly interacts with other pions, nucleons, kaons 
and resonances directly emitted on the freeze-out hypersurface, 
we focus on the interaction with the most abundant pion in this work. 
%Focusing on the pion cloud, we examine the self-energy 
%using the distribution function of emitted pions 
%and the forward $\pi\pi$ scattering amplitude in the elastic channels. 
%Assuming the thermal spectrum on the freeze-out hypersurface 
%and a simple time evolution within the free streaming, 
%we obtain the distribution after the ``freeze-out". 
The magnitude of the mean field interaction and the sign of its real part are determined using the results of 
a phenomenological analysis on the two-body $\pi\pi$ scattering amplitude\cite{Schenk,PI}, 
which fits the measured amplitude incorporating the constraint of the underlying chiral symmetry. 
%We find the momentum dependence of the mean field interaction 
%especially due to the $\rho$ meson resonance peak. 
The presence of the $\rho$ meson resonance peak plays important roles in 
both the real and imaginary parts of the mean field interaction.

\begin{figure}[t]
\begin{minipage}[t]{0.5\hsize}
  \begin{center}
        \includegraphics[scale=0.3]{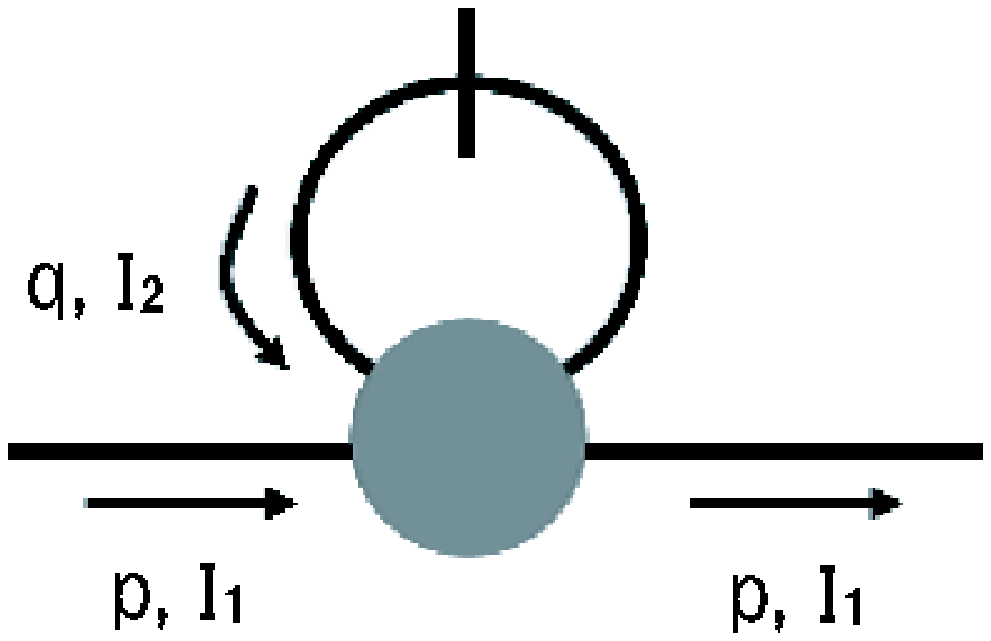}      
  \end{center}
  \caption{Mean field interaction with thermal excitations: 
In the pion cloud, an escaping pion picks up a thermal pion 
carrying the momentum $\bp^\prime$ and the isospin $I_2$. 
The gray blob represents the forward scattering amplitude in the vacuum. 
%The momentum of the medium particle, $\bm q$, is integrated out in Eq.(\ref{eq:sf}) 
%with the weight given by the distribution function $f(q)$. 
}
   \label{fig:diargam}
\end{minipage}
\begin{minipage}[t]{0.5\hsize}
  \begin{center}
        \includegraphics[scale=0.45]{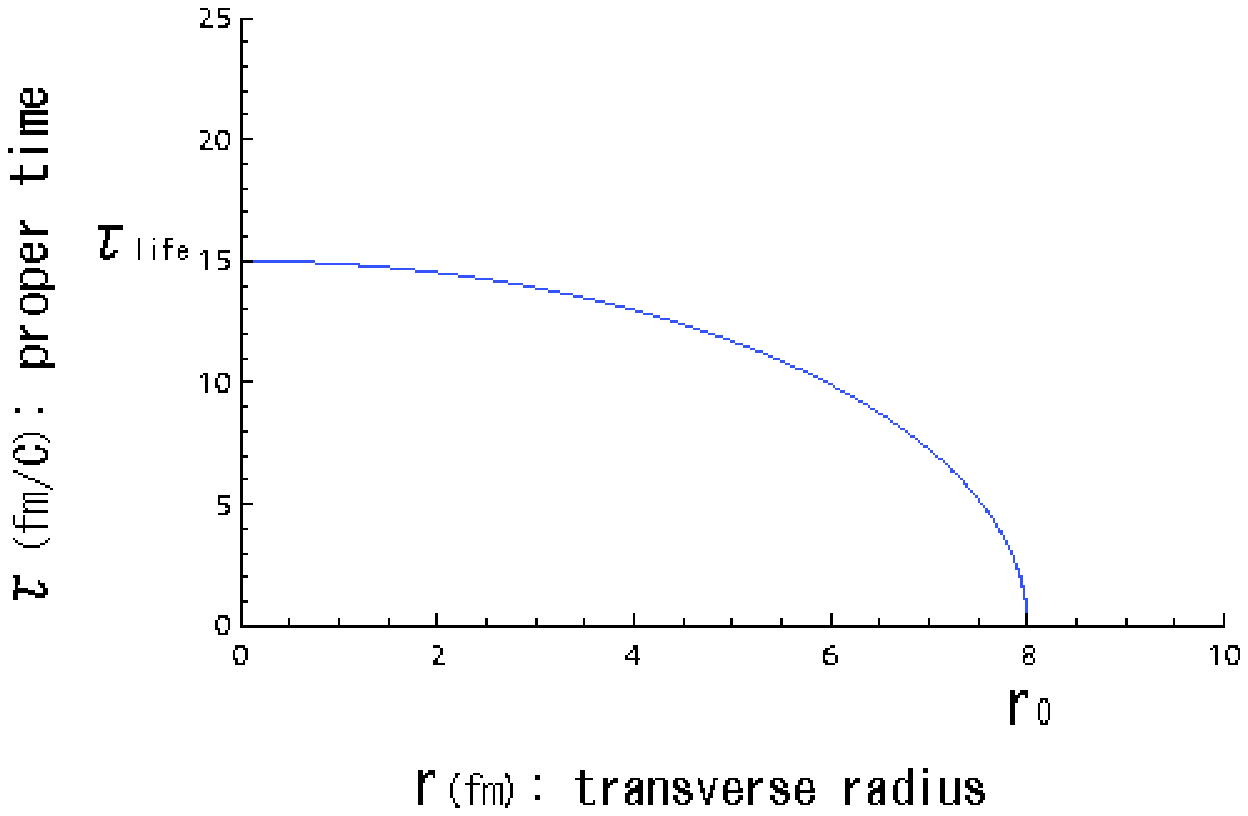}      
  \end{center}
  \caption{A model profile of the freeze-out proper time: 
On the transverse plane, the ``freeze-out" in the surface region precedes that in the interior. 
Thus, the freeze-out surface shrinks as time advances. }
   \label{fig:surface}
\end{minipage}
 \end{figure}

The modification of the dispersion relation is expressed with the self-energy $\Pi$: 
\begin{eqnarray}
E_{\bp}^2 = {\bm p ^2} + m^2 + \Pi(\bp) \ ,   \label{eq:disp}
\end{eqnarray}
where $\bp$ and $m$ are the momentum and mass of the pion escaping from the cloud, respectively. 
Since %, in the time evolution of matter generated in the ultrarelativistic heavy ion collision, 
the system becomes diluted in the freeze-out stage owing to the evaporation, 
%promoted by both the longitudinal and radial expansions, 
two-body scattering of thermal pions dominates the contribution to the pion self-energy. 
In Fig. \ref{fig:diargam}, we sketch the scattering with a thermal excitation, 
where the closed line indicates an excited medium pion carrying momentum $p^\prime$. 
Up to the two-body scattering of on-shell excitations, 
the diagram contains only a single thermal loop, 
and the gray blob represents the forward scattering amplitude in the vacuum. 
The pion self-energy $\Pi$ described in Fig. \ref{fig:diargam} is written as 
\begin{eqnarray}
\Pi^{I_1}(\bp) = - \sum_{I_2=0,\pm 1} \integ \frac{d^4p^\prime}{(2\pi)^4} \   
2\pi \delta(p^{\prime 2}-m^2) \ T^{I_1 I_2}(s) \  f^{I_2}(p^\prime) \ , \label{eq:sf0}
\end{eqnarray}
where the isospin component of the medium pion, $I_2$, is summed over -1, 0 and +1, 
and $s$ is the Mandelstam variable, 
that is, the squared center-of-mass energy of the two-body scattering defined by $s=(p+p^\prime)^2$. 
$T^{I_1 I_2}(s)$ and $f^{I_2}(p^\prime)$ are the two-body forward scattering amplitude in the vacuum 
and the distribution function of the medium particle 
carrying the momentum $p^\prime$ and the isospin $I_2$, respectively. 
The overall sign is up to the definition of the phase shift in $T^{I_1 I_2}(s)$; 
we take the minus sign so that we have a negative phase shift for repulsive scattering. 
The distribution function and the on-shell condition %for the momentum $p^\prime$ of the medium pion 
come from the thermal part of the pion propagator. 
Here, we suppose that the renormalization has already been achieved in Eq. (\ref{eq:sf0}), 
and input the physical pion mass, $m=139.5$ MeV, 
and the phenomenological amplitude into Eq. (\ref{eq:disp}). 
The integral with respect to the momentum $p^\prime$ does not 
pose further ultraviolet divergence owing to the distribution function.

Neglecting the small difference in the light quark masses, 
the scattering amplitude by the strong interaction 
is blind to the third component of the total isospin. 
Here, we assume that each species of pion is produced in the same amount 
in the RHIC experiment\cite{PtPHENIX,PtSTAR}, 
which we write as $f(p^\prime)=f^{I_2}(p^\prime)$ for $I_2=0,\pm1$. 
In the isospin symmetric limit, the right-hand side of Eq. (\ref{eq:sf0}) is then simplified to 
\begin{eqnarray}
\Pi(\bp) &=& -  \integ \frac{d^4p^\prime}{(2\pi)^4} \   
2\pi \delta(p^{\prime 2}-m^2) \ T(s) \  f(p^\prime)       \ ,   
 \label{eq:sf}\\
T(s) &=& 3 \left( \frac{1}{9} T_0 (s) + \frac{3}{9}T_1 (s) + \frac{5}{9}T_2 (s) \right)   
\ , \label{eq:Tave}
%T(s) &=&  \frac{1}{9} T^0 (s) + \frac{3}{9}T^1 (s) + \frac{5}{9}T^2 (s) \ . \label{eq:Tave}
\end{eqnarray}
where, owing to the isospin symmetry, 
the self-energy is independent of the isospin carried by an escaping pion. 
%which is described by the external line in Fig.\ref{fig:diargam}. 
The sum over the isospin component of the medium pion in Eq. (\ref{eq:sf0}) results in 
the overall multiplication factor attached to 
the isospin-averaged scattering amplitude $T(s)$ in Eq. (\ref{eq:Tave}). 
The subscripts on the right-hand side of Eq. (\ref{eq:Tave}) denote the total isospin channels. 
To evaluate the self-energy given by Eq. (\ref{eq:sf}), 
we need two distinct physical quantities, $T(s)$ and $f(p^\prime)$, 
which provide a microscopic foundation to the mean field interaction. 
%and incorporate the space-time geometry of the source. 

We first examine a simple time evolution of the distribution function $f(p^\prime)$. 
The dynamics of the expanding source is important in two respects. 
First, the density of pions, strolling in the vicinity of the source, 
becomes diluted owing to the evaporation, 
which diminishes the modification of the dispersion relation as time advances. 
In the time evolution of the matter created in the ultrarelativistic heavy ion collision, 
the evaporation is promoted by 
the transverse expansion and the approximately boost invariant strong longitudinal expansion. 
Second, the collectivity changes the typical scattering energy of pions compared with the case without it. 
%Second, the collectivity leads to the rather low energy scattering of pions, 
%since the relative momenta of them are typically smaller compared to the case without it. 
This effect reflects in the self-energy 
through the Mandelstam variable in the scattering amplitude, $T(s)$.

To reflect the space-time geometry of the source, we adopt a phenomenological extension to Eq. (\ref{eq:sf}). 
The distribution function is here assumed to depend on the space-time coordinate as well as the momentum. 
Assuming the free streaming of medium pions for the first approximation, 
the self-energy is then obtained by solving the free classical transport equation: 
\begin{numcases}
{}
\Pi(t, \bx, \bp) = -  \integ \frac{d^4p^\prime}{(2\pi)^4} \   
2\pi \delta(p^{\prime 2}-m^2) \ T(s) \  f(t, \bx,\bp^\prime)   \label{eq:sf10} \\
p^{\prime\mu}\partial_\mu f(t, \bx,\bp^\prime) = S(t, \bx,\bp^\prime) 
\label{eq:Bol}
\end{numcases}
where the self-energy $\Pi(t, \bx, \bp)$ depends on 
the momentum $\bp$ through the center-of-mass energy, $\sqrt{s}$. 
A source term $S(t, \bx,\bp^\prime) $ describes the emission at the freeze-out hypersurface. 
Solving Eq. (\ref{eq:Bol}) under the initial condition, $f(t=0,\bx,\bp^\prime) = 0$, 
we obtain the distribution of the pion after the emission.

To discuss the longitudinal boost invariance, 
it is convenient to introduce the kinematical variables defined by 
\begin{displaymath}
\left\{
\begin{array}{l}
Y = \frac{1}{2} \log \frac{E_{\bp} + p_z}{E_{\bp} - p_z}  \ , \\
m_\perp = \sqrt{ p_\perp^{2} + m^2 } \ , 
\end{array}
\right.  \ \ \ \ {\rm and}  \ \ \ \ 
\left\{
\begin{array}{l}
\eta = \frac{1}{2} \log \frac{t + z}{t - z} \ , \\
\tau = \sqrt{ t^2 - z^2 } \ , 
\end{array}
\right. 
\end{displaymath}
where $p_\perp$ is the magnitude of the transverse momentum, $\bp_\perp=(p_x,p_y)$. 
%The rapidity and the space-time rapidity have the additive property for the Lorentz boost in the longitudinal direction. 
Using these variables, we rewrite Eq. (\ref{eq:Bol}) to exhibit the manifest symmetries as 
\begin{eqnarray}
%&&
\left( \frac{\partial }{\partial \tau} - \frac{1}{\tau} \tanh\xi \frac{\partial }{\partial \xi} 
+ \frac{1}{m_\perp \cosh\xi} \ \bp_\perp \cdot \nabla_\perp \right) f(\tau,r,\bp_\perp,\omega,\xi)
= \frac{S(t,\bx,\bp)}{m_\perp \cosh\xi} \  ,
\label{eq:inv}
\end{eqnarray}
where we dropped the primes on the momenta. 
We have the two differences $\xi=Y-\eta$ and $\omega = \phi - \theta$, 
where the angles, $\theta$ and $\phi$, are the arguments of the transverse coordinate $\bx_\perp$ 
and the transverse momentum $\bp_\perp$, respectively. 
These differences are invariant under the Lorentz boost in the longitudinal direction and the rotation on the transverse plane, respectively. 
We find the manifest cylindrical symmetry on the left-hand side of Eq. (\ref{eq:inv}), 
noting that the inner product of the vectors on the transverse plane is given by 
the difference in the angles, $\bp_\perp \cdot \nabla_\perp f = |\bp_\perp| |\nabla_\perp f| \cos \omega$. 
Thus, the cylindrical symmetry of the distribution function is preserved under the time evolution, 
as long as the source function has the same symmetry in the form, $S(\tau,r,\bp_\perp,\omega,\xi)$.

Provided that the evaporating pions have the thermal spectrum on the freeze-out hypersurface, 
the source functions $S(t, \bx,\bp)$ in Eqs. (\ref{eq:inv}) and (\ref{eq:P1free}) are consistently given by 
\begin{eqnarray}
S(t, \bx, \bp) &=& \frac{1}{(2\pi)^3} 
\int f_{eq}(x^{\prime}, p)\  \delta^{4}(x-x^{\prime}) \ p^{ \nu} d\sigma_{\mu}(x^\prime) \ , \label{eq:source}  \\
f_{eq}(x, p)  &=& \frac{1}{\exp \{ (p^{ \nu} u_{\nu}-\mu)/T \} -1 }  \ , \label{eq:feq}
\end{eqnarray}
where $x^\prime$ is the emission point on the hypersurface, and $d\sigma_{\mu}(x^\prime)$ is the normal vector of it. 
The integral is carried out over the hypersurface owing to the delta function. 
The thermal distribution function $f_{eq}(x, p)$ is specified by %the isotropic thermal spectrum on the rest frame of the fluid cell 
the macro variables: temperature $T$, pion chemical potential $\mu$ and the flow vector $u^{\mu}$. 
The longitudinally boost invariant expansion is described using the Bjorken flow\cite{Bjo83}: 
\begin{eqnarray}
u^{\mu}(x) = \frac{1}{\sqrt{1-\bm v_\perp^2}} \left( \frac{t}{\tau}, \bm{v}_{\perp}, \frac{z}{\tau} \right) \ ,
\end{eqnarray}
where %$\tau$ is the longitudinal proper time, $\tau=\sqrt{t^2-z^2}$, and 
$\bm{v}_{\perp}$ is the transverse flow vector. 
Its magnitude depends on the transverse radial coordinate, $r$. 
The energy on the rest frame of the fluid cell is written with the flow vector as 
\begin{eqnarray}
p^{\nu} u_{\nu} = \frac{ m_{\perp} \cosh \xi - p_{\perp} v_{\perp} \cos \omega }
{ \sqrt{1-v_{\perp}^2} } \ ,
\end{eqnarray}
where $p_{\perp}$ and $ v_{\perp}$ are the magnitudes of the transverse momentum and the transverse flow velocity. 
Owing to the above expression, we find the boost invariance and rotational invariance 
of the source function in Eqs. (\ref{eq:source}) and (\ref{eq:feq}). 
The normal vector of the hypersurface is given by 
\begin{eqnarray}
d\sigma^{\mu}(x) = \tau_f \left( 
\cosh\eta,  \cosh\eta \frac{\partial t_f(r,z)}{\partial r} \bm{e}_\perp, \sinh\eta
\right ) r dr d\theta d\eta
\ , \label{eq:Vnorm}
\end{eqnarray}
where the freeze-out time, $t_f(r,z)=\sqrt{\tau_f^2(r)+z^2}$, is specified by 
the $r$-dependent freeze-out proper time $\tau_f(r)$ given below. 
On the transverse plane, 
the ``freeze-out" may proceed from the surface region to the deep interior 
due to the finite source volume and the finite sound velocity.

Motivated by the hydrodynamical simulations\cite{HT02,KH03}, 
we have used the following parameters and analytic functions in the numerical analysis. 
We take $T=130$ MeV and $\mu=30$ MeV for the temperature and the pion chemical potential, respectively. 
%and analytic forms of the radial flow $\bm v_\perp$ and the freeze-out proper time $\tau_f(r)$. 
The profile of the radial flow is approximated using a linear function of the radial coordinate $r$, $|\bm v_\perp| = 0.06\hspace{0.1cm} r \; c^{-1}$, 
and the freeze-out proper time using an ellipsoidal curve, $\tau_f(r) = \tau_{\rm life} \sqrt{ 1-r^2/r_0^2 }$. 
The two radii, $\tau_{\rm life}$ and $r_0$, are roughly the lifetime of the source 
and the radius of the cylinder soon after the nucleus collision, respectively(see Fig. \ref{fig:surface}). 
We have chosen $\tau_{\rm life}=15$ fm/$c$ and $r_0=8$ fm.

Concerning the classical transport equation in Eq. (\ref{eq:Bol}), 
we should comment on the following two points. 
First, the source term introduced to describe the emission on the freeze-out hypersurface 
can be negative at the timelike part of the hypersurface, owing to the longstanding issue of the Cooper-Frye formula. 
Its contribution induces a negative value of the distribution function $f(t,\bx,\bp)$. 
To cope with this issue, we have confirmed to what extent it could reflect in our results. 
If we find the negative distribution, we cut off its contribution to the integral in Eq. (\ref{eq:sf10}). 
Comparing the results obtained with and without the negative distribution, 
we observed that the self-energy changes within 2\%, and the final results, 
the Gaussian radius parameters in Fig. \ref{fig:R} and the transverse spectrum in Fig. \ref{fig:spec}, change by not more than 1\%. 
We consider that the integral in Eq. (\ref{eq:sf10}) with respect to the configuration of medium pion smears the negative contribution. 
The negative contribution is negligibly small, 
at least within our choice of the parameters and the profile of the freeze-out hypersurface specified above. 
%at least as long as we choose the parameters and the profile of the hypersurface specified above. 
Although the longstanding issue of the Cooper-Frye formula still remains, 
%on the other hand, 
our results are barely affected.

Second, to treat the motion of medium pions self-consistent to the effects of the mean field interaction, 
it will be necessary to incorporate the Vlasov term and the collision term into the transport equation in Eq. (\ref{eq:Bol}). 
They are described using the self-energy $\Pi(t, \bx, \bp)$, which 
couples the transport equation (\ref{eq:Bol}) to the expression of the self-energy in Eq. (\ref{eq:sf10}). 
However, owing to the following discussions, 
we again deduce that the small deviations in the configuration of medium pions are smeared in the integral in Eq. (\ref{eq:sf10}), 
as is observed for the contribution of negative distribution.

The Vlasov term describes the variation of the distribution due to the momentum change, 
or the deviation of the classical trajectory from the free motion, by the external force. 
We find in the equation of motion (\ref{eq:cano}) that 
the force is proportional to the gradient of the real part of the self-energy. 
The validity to neglecting the Vlasov term is not obvious {\it a priori}. 
%However, owing to our results obtained assuming the free motion, 
%we expect that this treatment is a good first approximation. 
Nevertheless, 
in the computation of the single-pion spectrum shown in Fig. \ref{fig:spec}, 
we find that the shift of the transverse momentum by the mean field interaction is less than 20 MeV. 
Provided that the momentum shift is caused in the time scale on the order of the duration, $\tau_{\rm life}$, 
the deviation of the classical trajectory from the free motion is estimated to be 
$| \Delta \bx | \sim \frac{ | \Delta \bp | }{E_\bp} \tau_{\rm life} \lesssim 1 \ {\rm fm}$. 
The momentum shift and the deviation of the trajectory are 
almost an order smaller than the magnitude of the momentum and the extension of the source region, $r_0$, respectively.

\begin{figure}[t]
  \begin{center}  
  \includegraphics[scale=0.48]{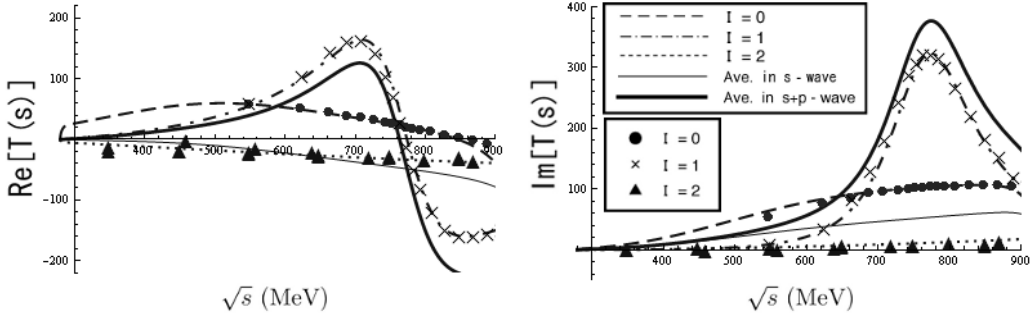}  
  \end{center}
\caption{Elastic forward $\pi\pi$ scattering amplitude $T^I(s,0)$ in the s-wave and p-wave: 
The panels on the left and right show the real and imaginary parts of the amplitude, respectively. 
The horizontal axes show the center-of-mass energy, $\sqrt{s}$, of the two-body scattering. 
Curves show the amplitudes in each isospin channel, $T^0$, $T^1$ and $T^2$, 
and their taken with the degeneracy factor for the isospin multiplet, as shown in Eq. (\ref{eq:Tave}). 
We find a narrow peak in the imaginary part of the $I=1$ channel around $\sqrt{s}\simeq 770$ MeV, 
located on the mass of $\rho$ meson resonance. 
The experimental data\cite{PiPiex} are indicated by markers for the isospin channels. 
}
\label{fig:amp}
\end{figure}

The collision term describes the loss and gain of the pions in the phase space volume element. 
The optical theorem relates the imaginary part of the forward scattering amplitude $T(s)$ 
to the total cross section of the binary collision, $\pi\pi\rightarrow\{{\rm anything}\}$. 
Thus, the loss of the pions distributed in the interval, $\bx \sim \bx+\Delta\bx$ and $\bp \sim \bp+\Delta\bp$, at $t$ 
is described by the imaginary part of the self-energy as 
${\rm Im} \left[\Pi(t, \bx, \bp)\right] f(t, \bx, \bp)$ up to the normalization. 
In Fig. \ref{fig:amp}, we find that the imaginary part of $T(s)$ is dominated by $\rho$ meson resonance peak. 
Therefore, the loss is mainly attributed to the formation of $\rho$ meson resonance, 
$\pi(p)\pi(p^\prime)\rightarrow\rho(p+p\prime)$.\footnote{
The absorption by the mean field interaction is thus microscopically originated in this reaction.} 
The gain is dominated by the inverse reaction, that is, the decay of $\rho$ meson resonance 
$\rho(p+p\prime)\rightarrow\pi(p_1)\pi(p_2)$, which has the almost 100\% branching ratio. 
Because this reaction, $\pi(p)\pi(p^\prime)\rightarrow\rho(p+p\prime)\rightarrow\pi(p_1)\pi(p_2)$, 
appears in the p-wave scattering, 
the differential cross section of this process is proportional to the square of the Legendre function, 
$|P_{\ell=1}(\cos\theta)|^2=\cos^2\theta$, where $\theta$ is the scattering angle in the center-of-mass frame. 
The momenta in the final state are localized in the forward direction: 
therefore, we have $p\sim p_1$ and $p^\prime\sim p_2$, 
or $p\sim p_2$ and $p^\prime\sim p_1$, owing to the momentum conservation. 
Although the finite width of the angular distribution in the final state causes 
a small variation of the distribution function, this effect is smeared in the integral in Eq. (\ref{eq:sf10}). 
%Therefore, the effect of the collision term in the freeze-out stage will not 
%change the self-energy $\Pi(t, \bx, \bp)$ drastically, 
%although the two-particle correlation is more sensitive to the variation of $f(t, \bx, \bp)$. 
As far as we use the distribution of the free streaming pions to obtain the self-energy, 
we expect that the effects of the Vlasov term and collision term are, in practice, not relevant to the purpose of this work. 
However, a self-consistent framework with these terms will be necessary 
to describe possibly more efficient effects of the mean field interaction in the denser system.

%%%%%%%%%%%%%%%%%%%%%%%%%%%%%%%%%%%%

Next, we show the momentum dependence of the forward $\pi\pi$ scattering amplitude in the vacuum. 
Pion is known as an approximate Nambu-Goldstone boson, 
which has been studied on the basis of the underlying chiral symmetry, and its breaking. 
%Owing to PCAC hypothesis, the low-energy theorem predicts that the s-wave scattering lengths 
%are proportional to the square of pion mass $m_{\pi}$\cite{Wei66}. 
%Because it is the lightest hadrons, 
%their two-body reaction is dominated by elastic scattering below the threshold of two kaons, $\sqrt{s}\sim 1$GeV. 
Here, we use a phenomenological analysis of the $\pi\pi$ scattering amplitude in the s-wave and p-wave. 
It was achieved on the basis of the Roy equation\cite{PI}, 
which is the manifestation of the fundamental properties of the scattering amplitude: 
analyticity, unitarity and crossing symmetry. 
This analysis incorporates the constraint by the chiral symmetry, 
and precisely reproduces the behavior of the $\rho$ meson resonance peak.

In each total isospin channel, 
the scattering amplitude $T^I(s,t)$ is decomposed into the partial waves as 
\begin{eqnarray}
T^I(s,t) &=& 32\pi \sum_{\ell} (2\ell +1) 
P_{\ell}\left( 1+\frac{2t}{s-4m_{\pi}^2} \right) t_{\ell}^I(s) \ , \\
t_{\ell}^I(s) \;\; 
&=& \frac{1}{2i\sigma(s)} \left(  e^{2i\delta_{\ell}^I(s)} -1 \right) \ , \\
\sigma(s) \;\; &=& \sqrt{ 1 - \frac{4m_{\pi}^2}{s} } \ ,%\;\;=\;\; 2 \frac{ p_{cm} }{ E_{cm} }
\end{eqnarray}
where the Mandelstam variables, $s$ and $t$, are related to the energy and the scattering angle in the center-of-mass frame. 
Below the threshold of two kaons, $\sqrt{s_{\pi\pi}}\sim 1$ GeV, we focus on the elastic scatterings without the inelasticity. 
The analysis provides the phase shift $\delta_{\ell}^I(s)$ in the s-wave and p-wave. %where the elasticity is identical to one, $\eta_{\ell}^I(s)=1$. 

We show the forward scattering amplitude $T^I(s,0)$ in Fig. \ref{fig:amp}. 
The panels on the left and right show the real and imaginary parts of the amplitude, respectively. 
The horizontal axes show the center-of-mass energy of the two-body scattering. 
Curves display the amplitudes in each isospin channel, $T^0$, $T^1$ and $T^2$, 
and their average taken with the degeneracy factor for the isospin multiplet as shown in Eq. (\ref{eq:Tave}). 
The maximum value in the $I=1$ channel is three times larger than those in the other two channels 
owing to the degeneracy factor for the angular momentum, $\ell$.

%As is shown by the low energy theorem given in eq.(\ref{eq:achiral}), 
In the s-wave, 
the real part of the amplitude in the $I=0$ channel has an attractive regime at low energy, 
while that in the $I=2$ channel exhibits weak repulsion. 
The phase shift passes $\pi/2$ at the intercept of the horizontal axis on the left panel, 
or the center of the peak on the right panel. 
The amplitude in the $I=0$ channel shows a slow variation against the increase in $\sqrt s$, 
and the imaginary part of it has a broaden peak. 
The energy dependence in the $I=2$ channel is much weaker, 
and the phase shift even does not pass $\pi/2$ in the range shown in these plots. 
On the other hand, the amplitude in the $I=1$ channel has a strong energy dependence. 
Owing to the optical theorem, 
the narrow peak in the imaginary part is related to the peak in the cross section, 
and its location at $\sqrt{s}\simeq 770$ MeV is identified to the mass of $\rho$ meson resonance. 
While the average in the s-wave scattering displayed with a thin solid line exhibits a weak repulsion, 
the contribution of p-wave scattering overwhelms it to provide 
an attraction below the mass of $\rho$ meson, as indicated by the thick solid line. 
On the right panel, 
the thick solid line for the average is close to the dash-dotted line for the $I=1$ channel. 
The absorptive effect described by the imaginary part is dominated by the formation of the $\rho$ meson resonance in the $I=1$ channel.

\begin{figure}[t]
  \begin{center}  
     \includegraphics[scale=0.36]{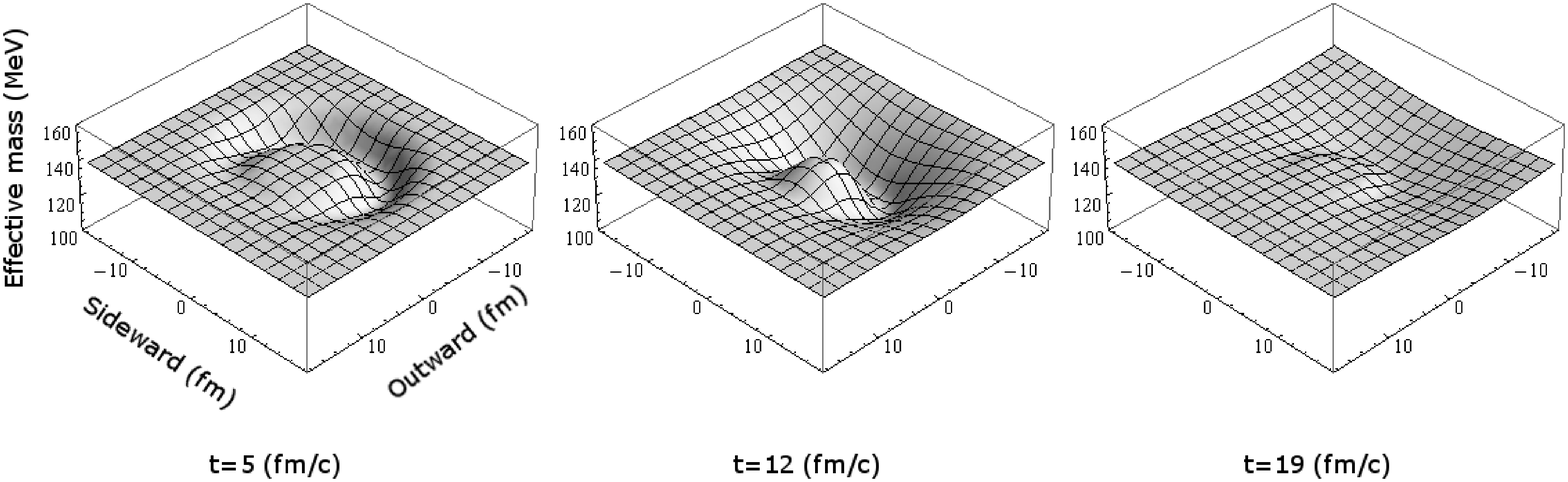}  
  \end{center}
\caption{
Time evolution of the effective pion mass defined by the real part of the self-energy in Eq. (\ref{eq:meff}): 
We show the two-dimensional spatial profiles of the effective pion mass at the transverse momentum, $p_\perp=150$ MeV. 
In this regime, the mean field interaction acts on the HBT images most effectively. 
%We show the case with the vanishing longitudinal momentum, $p_z=0$, where the experimental analysis has been carried out, 
%and the zero angle of the transverse momentum, $\theta=0$, without loss of generality owing to the rotational symmetry. 
We find the effects of the attractive mean field interaction in the vicinity of the freeze-out hypersurface, 
in the exterior of which any interactions are assumed to be absent in the hydrodynamical modeling. 
The sharp boundary is smeared by the mean field interaction, and the smeared boundary shrinks as time advances. 
We do not find the effect of the repulsive mean field interaction in these plots. 
}
\label{fig:time}
  \begin{center}  
     \includegraphics[scale=0.36]{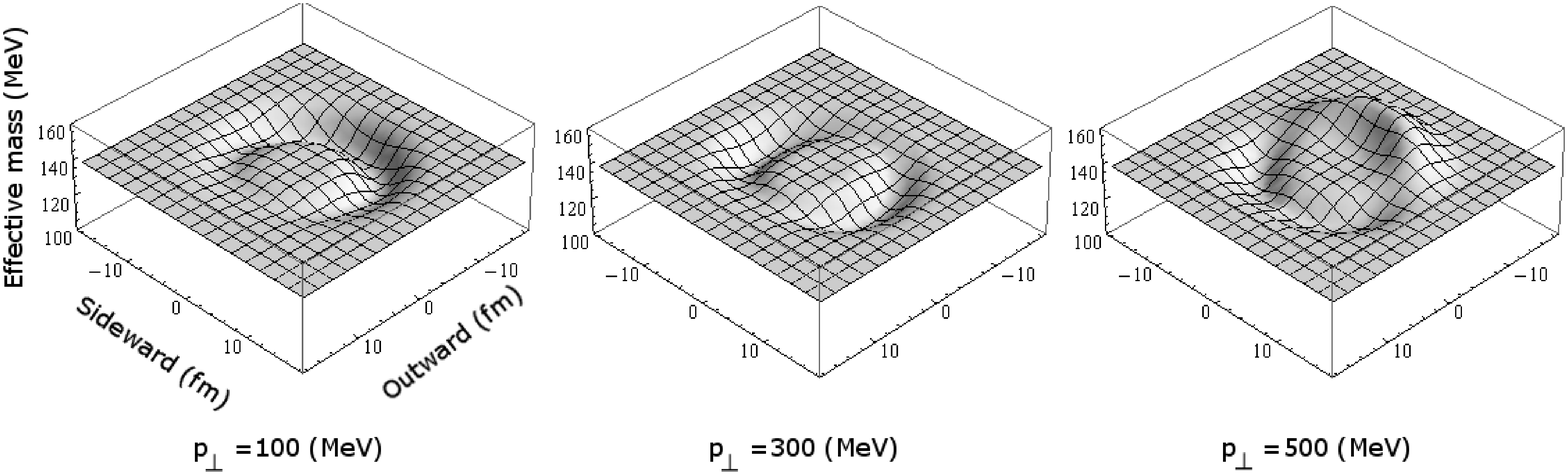}  
  \end{center}
\caption{
Momentum dependence of the effective pion mass: 
These panels exhibit the transverse momentum dependence of the effective pion mass at the moment, $t=5$ fm/$c$. 
%The longitudinal coordinate $z$, the longitudinal momentum $p_z$ and the angle of the transverse momentum $\theta$ 
%are chosen as those in Fig.\ref{fig:time}. 
At $p_\perp=500$ MeV, we find effect of the repulsive mean field interaction in the backside region of the source, 
where the center-of-mass energy of $\pi\pi$ scattering is typically large owing to the collective expansion of the medium. 
}
\label{fig:mom}
\end{figure}

\begin{figure}[t]
  \begin{center}
        \includegraphics[scale=0.33]{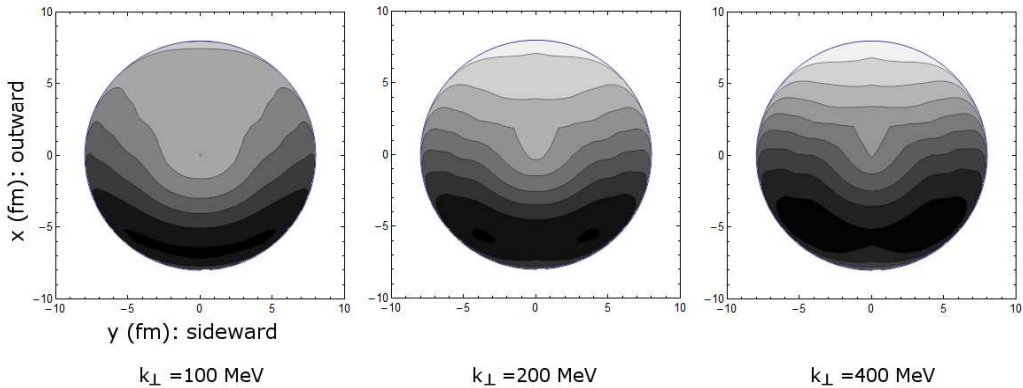}      
  \end{center}
  \caption{
Contour plot of the attenuation factor for the pion emitted at the position $(x,y)$ 
and detected with the asymptotic momentum $k_\perp$: 
The attenuation becomes stronger as the color becomes darker. 
As given by Eq. (\ref{eq:att}), 
the integral of the imaginary part of the self-energy is carried out along the world line, 
which is computed under the effect of the real part. 
Each position on the panel is the emission point located on the freeze-out hypersurface viewed on the transverse plane. 
Note that the emission at each time is plotted on the same panel. 
}
\label{fig:damp}
\end{figure}

Carrying out the convolutional integral in Eq. (\ref{eq:sf10}), 
we obtain the phenomenological self-energy. 
To illustrate the effect of the real part of the self-energy, 
we show the effective pion mass, which is defined by 
\begin{eqnarray}
m_{\rm eff}(t,\bx,\bp)=\sqrt{m^2 + {\rm Re}\Pi(t,\bx,\bp)} \ .
\label{eq:meff}
\end{eqnarray}
In Figs. \ref{fig:time} and \ref{fig:mom}, we show the spatial profiles of the effective pion mass 
with its time dependence and momentum dependence, respectively. 
Note that we focus on the central rapidity, $p_z=0$, 
and that the argument of the transverse momentum $\bp_\perp$ carried by the escaping pion is fixed at $\phi=0$ without loss of generality.

In Fig. \ref{fig:time}, we show the time dependence of the effects of the mean field interaction, 
where time advances from left to right, and the transverse momentum is fixed at $p_\perp=150$ MeV. 
We find the reduction of the effective pion mass in the vicinity of the freeze-out hypersurface at each moment. 
Note that, since we do not consider the interactions in the interior of the hypersurface, 
we have a flat region around the origin with $m_{\rm eff}\simeq 139$ MeV, which shrinks as the ``freeze-out" proceeds. 
The reduction of mass is the effect of the attractive mean field interaction 
obtained on the basis of the microscopic two-body scattering at low energy. 
In the microscopic view, 
we observe that the contribution of the attractive channel exceeds that of the repulsive channel at $p_\perp=150$ MeV. 
The effect of the mean field interaction diminishes rapidly as the emitted pion recedes from the freeze-out hypersurface.

In Fig. \ref{fig:mom}, 
we show the transverse momentum dependence of the effective pion mass at the moment, $t=5$ fm/$c$. 
On the rightmost panel, 
we find the enhancement of the effective pion mass on the backside region,\footnote{
With respect to the origin set on the center of the nucleus collision,
the detector side indicates the source region in the direction of the averaged momentum of the pair. 
The backside is the other half of the source region. } 
which is the effect of the repulsive mean field interaction. 
Owing to the repulsive $\pi\pi$ scattering above the $\rho$ meson mass, 
the mean field interaction can be repulsive. 
Although one may consider that the attractive $\pi\pi$ scattering at low energy dominates 
the contribution to the convolutional integral in Eq. (\ref{eq:sf10}), 
the repulsion in the higher regime exceeds the attraction owing to the effect of the collective expansion. 
Because the pion emitted in the backside region penetrates into the interior of the source against the collective motion\footnote{
Note that we show the effective pion mass for $\phi=0$ in Figs. \ref{fig:time} and \ref{fig:mom}.}, 
the center-of-mass energy of the scattering with expanding medium is typically large 
compared with the case emitted on the detector side. 
%On the detector side, an escaping pion drifts collectively toward the asymptotic region. 
Owing to the large center-of-mass energy, the repulsive interaction observed on the rightmost panel appears 
at the momentum, $p_\perp=500$ MeV, which is less than the mass of $\rho$ meson resonance.

As mentioned below Eq. (\ref{eq:Vnorm}), 
we have a negative contribution of the distribution function $f(t,\bx,\bp^\prime)$ owing to the longstanding issue of the Cooper-Frye formula. 
Nevertheless, the enhancement of the effective mass is not an associate artifact of this issue. 
We have confirmed that the negative distribution results in less than 2\% variation of the self-energy.

In Fig. \ref{fig:damp}, we show the attenuation factor given by Eq. (\ref{eq:att}), 
which is obtained by integrating the imaginary part of the self-energy along the world line. 
We have computed the world line using the classical equation of motion (\ref{eq:cano}) 
with the real part of the self-energy. 
A set of boundary conditions is imposed by the emission point $(x,y)$ and the asymptotic transverse momentum $k_\perp$ indicated below each panel. 
The attenuation factor is displayed using the contour plot with the darker color for the stronger attenuation. 
Each position on the panel $(x,y)$ is the emission point located on the freeze-out hypersurface viewed on the transverse plane: 
therefore, the emission at each time is plotted on the same panel. 
Note that the emission point $(x,y)$ is not the apparent spatial coordinate $(x\prime,y^\prime)$ defined by Eq. (\ref{eq:shiftxy}), 
but the original coordinate before being transformed.

We find that the attenuation is strong for the pions emitted in the backside region, 
because the path length to escape the interaction range is large for them 
compared with the case emitted in the surface region on the detector side. 
This feature is common to the cases at different momenta. 
Comparison of the left two panels at $k_\perp=100$ MeV and $k_\perp=200$ MeV shows that 
the attenuation becomes stronger at $k_\perp=200$ MeV as the transverse momentum becomes larger. 
This is because the imaginary part of the scattering amplitude shown in Fig. \ref{fig:amp} 
is an increasing function of $\sqrt{s}$ below the $\rho$ meson mass. 
On the other hand, we find that the profiles of the attenuation factor at $k_\perp=200$ MeV and $k_\perp=400$ MeV 
are similar to each other almost independent of the transverse momentum. 
Since the imaginary part of the scattering amplitude is dominated by the $\rho$ meson peak, 
the convolutional integral in Eq. (\ref{eq:sf10}) converges 
once the center-of-mass energy $\sqrt{s}$ goes beyond the $\rho$ meson peak.

\subsection{Modification of the Gaussian radius parameters} \label{subsec:Rhbt}

\begin{figure}[t]
  \begin{center}  
     \includegraphics[scale=0.33]{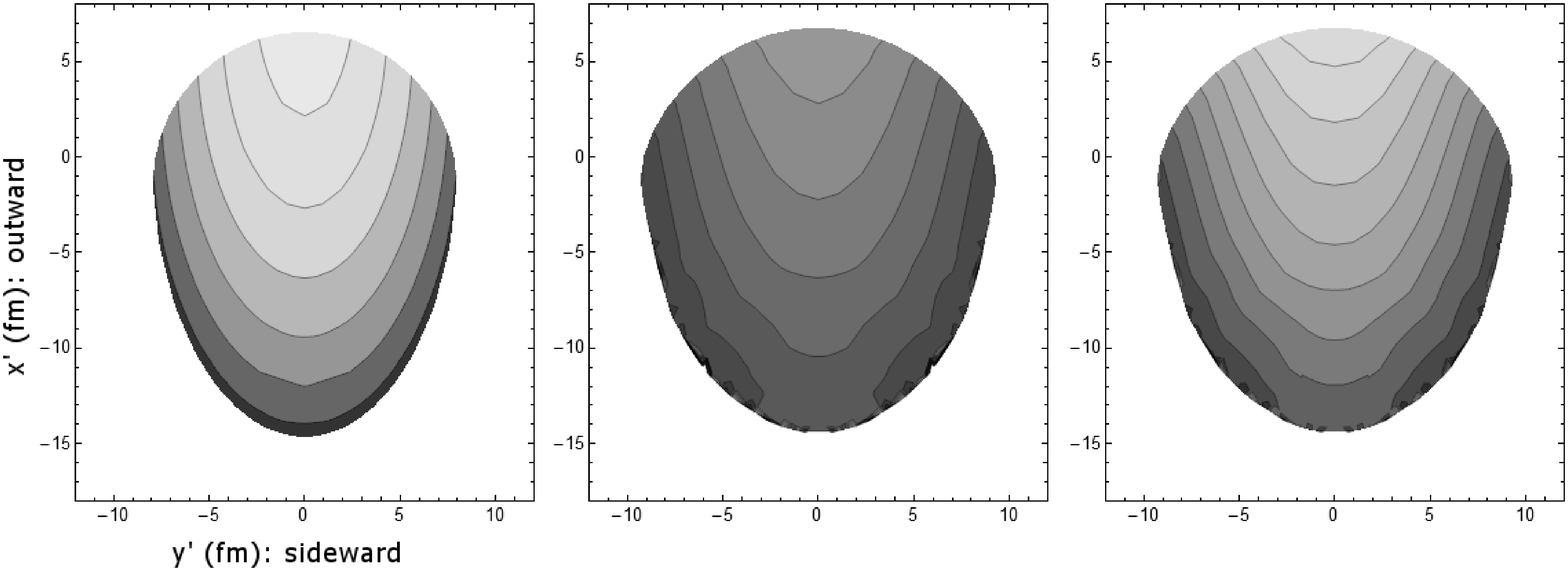}  
  \end{center}
\caption{
Profiles of the distorted images viewed on the apparent coordinate system at $k_\perp=200$ MeV: 
These panels show the contour plots of the effective distribution function in Eq. (\ref{eq:reff}). 
The location on each panel is the apparent emission point $(x\prime, y^\prime)$ 
transformed by Eq. (\ref{eq:att}) from the original coordinate $(x,y)$. 
The color becomes lighter as the distribution increases. 
The leftmost panel shows the free case, in which the image is elongated in the outward direction 
owing to the temporal structure of the source, although we have the cylindrical geometry for the spatial structure. 
We show the effects of the real part of the potential on the central panel, 
and the effects of both the real and imaginary parts on the rightmost panel. 
}
\label{fig:prof200}
\vspace{0.7cm}
  \begin{center}  
     \includegraphics[scale=0.33]{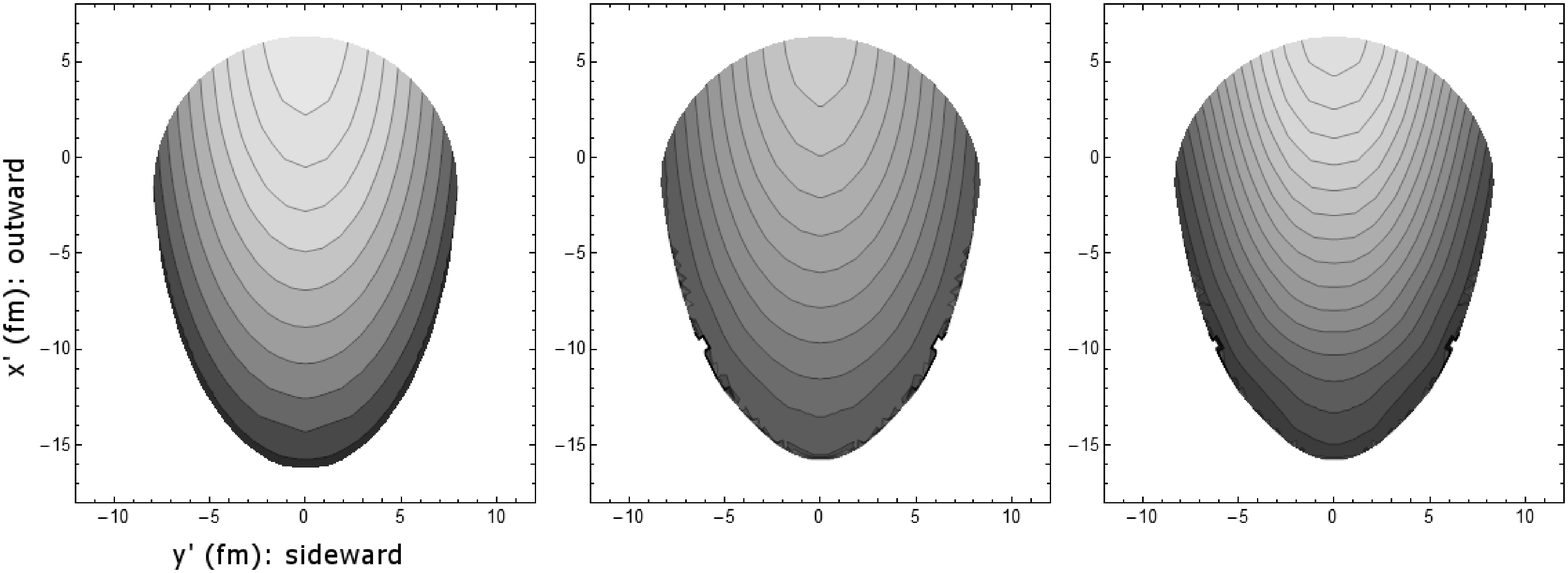}  
  \end{center}
\caption{
Profiles of the distorted images viewed on the apparent coordinate system at $k_\perp=400$ MeV: 
See the description in Fig. \ref{fig:prof200}. 
}
\label{fig:prof400}
\end{figure}

\begin{figure}[t]
  \begin{center}
        \includegraphics[scale=0.5]{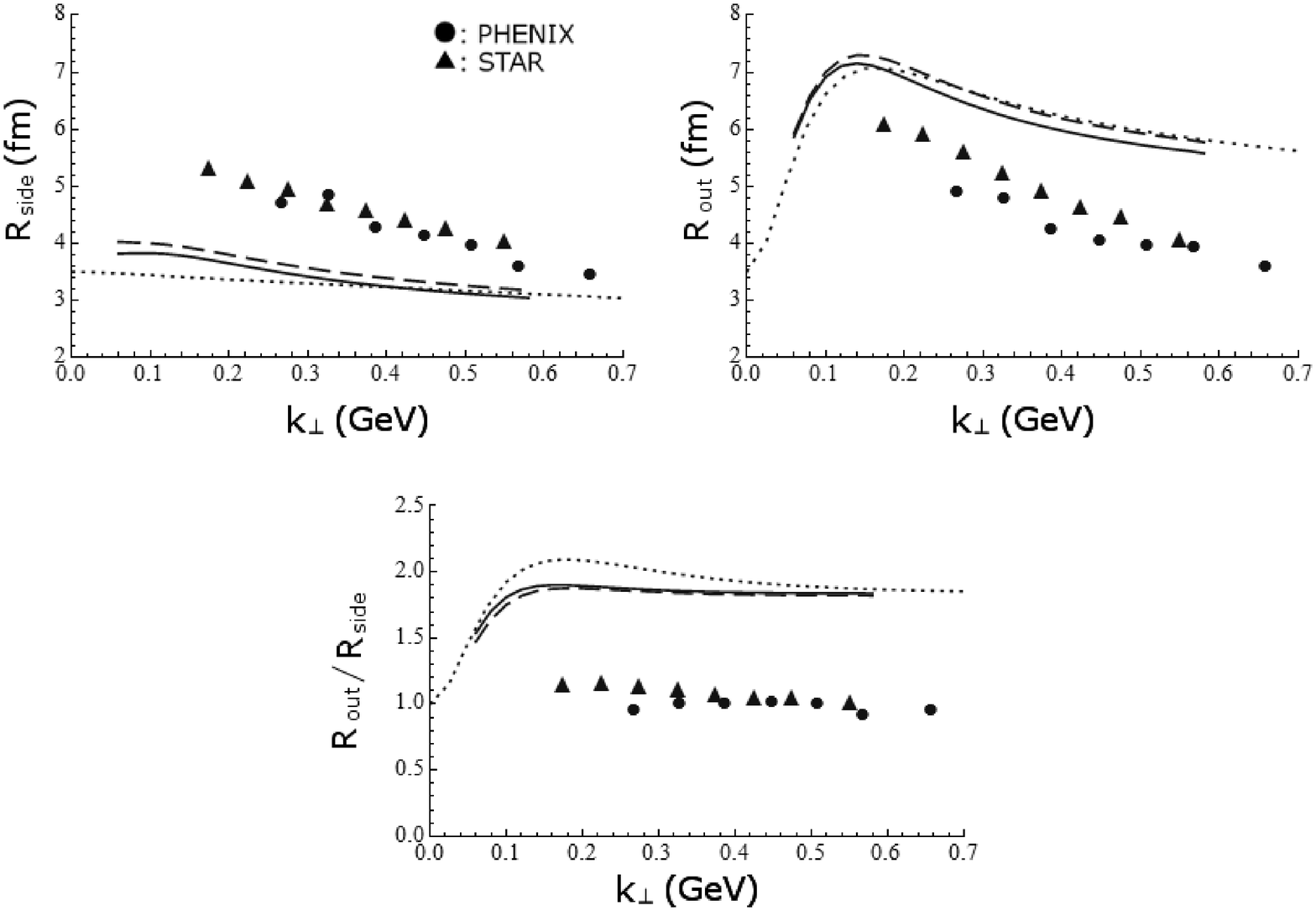}      
  \end{center}
  \caption{Gaussian radius parameters, $R\side$, $R\out$ and their ratio $R\out/R\side$: 
The dotted curve in each panel indicates the case without the mean field interaction. 
While the dashed curve indicates the effect of only the real part of the mean field interaction, 
the solid curve indicates the effect of both the real and imaginary parts. 
The effects of the mean field interaction are effective below $k_\perp \lesssim 0.5$ GeV. 
The experimental data measured in Au+Au collision at $\sqrt{s_{NN}}=200$ GeV 
are indicated by the circles and triangles\cite{PHENIX04,STAR05}.
}
   \label{fig:R}
\end{figure}

We investigate how and to what extent the dynamical mean field interaction 
can be effective to the observables measured in the ultrarelativistic heavy ion collision. 
Using the phenomenological mean field potential, or the self-energy, obtained in \S \ref{subsec:meanfield}, 
we show the distorted profiles of the HBT images, as in the static case examined in \S \ref{sec:prof}. 
To make these findings more quantitative, 
we show the effects on the conventionally analyzed Gaussian radius parameters. 
%The one-body potential describes the interactions in the exterior of the freeze-out hypersurface: 
%therefore, the complicated interaction with the matter is not included in our model. 
The phase shift is evaluated here by calculating the relativistic one-particle classical action as described in Appendix A, 
while we have performed nonrelativistic analysis in \S \ref{sec:prof}.

In Figs. \ref{fig:prof200} and \ref{fig:prof400}, we show the profiles of the distorted images viewed on the apparent coordinate system. 
Each panel shows the contour plot of the effective distribution function defined in Eq. (\ref{eq:reff}), as in Figs. \ref{fig:cntr100} and \ref{fig:cntr150}. 
The results shown in \S \ref{sec:prof} for the static source is improved here 
by incorporating the time dependence of the freeze-out hypersurface and using the phenomenological mean field interaction.  
For the dynamical source, we find the apparent deformation of the image even in the free cases 
as displayed on the leftmost panels in each figure. 
The images are elongated in the outward direction owing to the temporal structure of the source, 
although we have the cylindrical geometry for the spatial structure. 
Only the reflection symmetry, with respect to the outward axis, 
is preserved owing to the preferred orientation in which the pion pair is detected. 
If we analyze a pair carrying averaged momentum oriented in the other direction, %carrying the averaged momentum oriented in the other direction, 
we find source images obtained by rotating those in Figs. \ref{fig:prof200} and \ref{fig:prof400}. 
%The rotational symmetry of the system is hidden. 

In the free case, the image is more strongly elongated at the higher momentum, $k_\perp=400$ MeV, 
since the apparent shift of the spatial coordinate in Eq. (\ref{eq:att}) is proportional to the particle velocity: 
actually, it is given by the distance, $\bm v_{ k} (t_0 - t _{\Omega} )$, 
where $t_0$ and $t _{\Omega}$ are the spatial coordinate dependent emission time 
and an arbitrarily chosen origin of the temporal coordinate, respectively. 
As we find from the intervals of curves, 
the slope of the image at $k_\perp=400$ MeV is much steeper than the slope at $k_\perp=200$ MeV. 
It indicates that, at the higher momentum, emissions concentrate on the front surface of the source region. 
This is the effect of the radial flow embedded in the thermal distribution on the freeze-out hypersurface, 
which has been observed in the hydrodynamical picture\cite{KH03}. 
Since the magnitude of the radial flow is large in the surface region, 
energetic pions are likely to be emitted in the thin surface region on the detector side, 
owing to the strong boost of their momenta.

The remaining four panels show the effects of the mean field interaction. 
The central panel in each figure shows the effect of the real part, 
and the rightmost panel shows the effects of both the real and imaginary parts. 
The effect of the real part stretches the image in the sideward direction at $k_\perp=200$ MeV, 
and is less effective at $k_\perp=400$ MeV. 
Comparing with what we observed in Figs. \ref{fig:cntr100} and \ref{fig:cntr150} with a schematic potential, 
we find that these effects are caused by the attractive mean field interaction, which is obtained in \S \ref{subsec:meanfield}. 
Comparison of the rightmost panel with the central panel shows that the slope along the outward axis is steeper on the rightmost panel, 
which indicates a large distribution on the detector side. 
The effect of the absorption cuts off the emission in the backside region, 
and the surviving particle tends to come from the region on the detector side. 
As we have mentioned below Eq. (\ref{eq:reff}), 
the normalization of the effective distribution function is preserved to be one at each momentum 
even in the presence of the absorption, owing to the same absorptive effect on the single-particle spectrum. 
Therefore, the effective distribution function shows the relative distribution %on the coordinate space. 
of the emission points of the surviving particles.

To investigate these effects more quantitatively, 
we show the Gaussian parameters $R\out$, $R\side$ and their ratio $R\out/R\side$ in Fig. \ref{fig:R}. 
These analyses are based on the formulae given in Eqs. (\ref{eq:Rfree}), (\ref{eq:COORint}) and (\ref{eq:EXPint}). 
%and the definition of the deviation $\tilde{x}$ and the expectation value $\langle \mathscr{O} \rangle$ 
%are given in Eq.(\ref{eq:COORint}) and Eq.(\ref{eq:EXPint}), respectively. 
The horizontal axis in each plot is the magnitude of the asymptotic transverse momentum $k_\perp$ carried by a pion pair. 
The dotted curves display the free cases. 
They are the Gaussian fitting parameters of the distribution at the ``freeze-out" 
specified by Eqs. (\ref{eq:source}) and (\ref{eq:feq}), 
and qualitatively reproduce the results obtained by the hydrodynamical simulations\cite{LPSW}. 
While the Gaussian parameter for the sideward extension, $R\side$, measures the bare spatial extension, 
that of the outward extension, $R\out$, is enhanced by the temporal structure. 
They have exactly the same value at $k_\perp=0$, since we have no coordinate shift in Eq. (\ref{eq:Rfree}) for vanishing velocity, $v_\perp=0$. 
The filled triangles and circles indicate the data measured 
in the central Au+Au collision at $\sqrt{s_{NN}}=200$ GeV\cite{PHENIX04,STAR05}.

We show the effects of the mean field interaction with dashed curves and solid ones. 
The dashed curves indicate the effect of the real part of the mean field potential, 
and the solid ones the cooperative effects of the real and imaginary parts. 
We find that 
the sideward radius $R\side$ is increased by the effect of the real part, 
and that it acts more effectively in the low-momentum regime. 
This behavior qualitatively agrees with what we observed in Figs. \ref{fig:prof200} and \ref{fig:prof400}, 
and also in Figs. \ref{fig:cntr100} and \ref{fig:cntr150} with a schematic attractive potential. 
%This effect of the real part improves the deviation from the data 
%seen in the case without the mean field interaction. 
We find that the deviation from the data in the sideward radius is improved by the phenomenological mean field interaction. 
However, the pion density in the vicinity of the freeze-out hypersurface 
is not sufficiently large to solve the puzzle completely.

The absorption unexpectedly acts to reduce the sideward radius $R\side$, 
contrary to the findings in Figs. \ref{fig:cntr100} and \ref{fig:cntr150}. 
There, we have argued that the extension in the sideward is stretched 
by cutting off the emissions in the interior and backside region of the source. 
%This conflict manly comes from twofold. 
This conflict is due to two main reasons. 
One reason is that our mean field model incorporates the interactions 
only among the emitted pions, without the mean field interaction by the thermalized matter. 
If we incorporate the latter interaction, 
the absorption by the matter acts to increase the sideward radius as we have already discussed. 
This effect could be examined in the consistent description of the mean field interaction throughout the dynamics of the matter.

The other reason is found in the profiles of the attenuation factor shown in Fig. \ref{fig:damp}. 
Focusing on the variation of the magnitude along the line parallel to the outward axis, 
we find that the attenuation factor becomes larger as the distance from the sideward axis increases. 
This behavior is clearly seen at $k_\perp=100\ {\rm MeV}$, and is less pronounced at $k_\perp=200\ {\rm MeV}$ and $k_\perp=400\ {\rm MeV}$. 
Such a profile of the attenuation factor leaves a large distribution in the interior of the source compared with that in the peripheral region, 
and it obviously results in the reduction of the sideward radius. 
Recall that the imaginary part of the convolutional integral in Eq. (\ref{eq:sf10}) is dominantly contributed by the $\rho$ meson peak, 
and that it has little contribution from the high-energy scatterings beyond the range of the peak. 
Therefore, the integral monotonically increases as more energetic scatterings contribute to it, 
unless the scattering energy goes beyond the range of the peak. 
Because the energetic particle is likely to be emitted in the surface region owing to the effect of the radial flow, 
the integral has much contribution from the energetic scatterings in that region. 
These contributions cause the observed large attenuation factor in the peripheral of the source region at $k_\perp=100\ {\rm MeV}$. 
As the momentum of the escaping pion, $k_\perp$, increases, 
the attenuation factor becomes large in the interior of the source, 
while it is convergent in the surface region with the scattering energy going beyond the $\rho$ meson peak. 
Thus, the attenuation factor depends on the sideward coordinate much weakly at at $k_\perp=200\ {\rm MeV}$ and $k_\perp=400\ {\rm MeV}$. 
Note that the attenuation factor of the pion emitted in the front surface region on the detector side is not large, 
since it drifts for the collective motion of the medium, and the center-of-mass energy is typically small. 

\if 0
the attenuation factor depends on the configuration of the medium pion much weakly, 
because the integral is convergent with respect to the scattering energy. 
This is why we find the almost constant attenuation factor along the constant $x$, at $k_\perp=200\ {\rm MeV}$ and $k_\perp=400\ {\rm MeV}$. 
Note that the attenuation factor of the pion emitted in the front surface region on the detector side is not large, 
since it drifts for the collective motion of the medium. Then, the center-of-mass energy is typically small. 
\fi

The effect of the absorption slightly reduces the outward radius $R\out$. 
This behavior qualitatively agrees with our intuition and the clear observations in Figs. \ref{fig:cntr100} and \ref{fig:cntr150}: 
that is, the emissions in the backside region are cut off by the absorption. 
The magnitude of the effect is, however, not so strong. 
It is largely because the mean field interaction by the matter is not incorporated, as mentioned above. 
Further investigations are needed to examine the opacity due to the mean field interaction. 
Owing to the interaction with the matter, 
the effect of the absorption would be enhanced 
and act to reduce the deviation in the outward radius.

Owing to the attractive interaction, 
the ratio, $R\out/R\side$, is improved by 15\% below 300 MeV. 
The magnitude of the effect is, however, quantitatively insufficient to resolve the deviation completely. 
By the hydrodynamical simulations, 
the ratio is overestimated owing to the overestimation of $R\out$, 
and the underestimation of $R\side$, in contrast. 
Thus, the effect of improving the deviation should act on the radii in a different way. 
%some effects to decrease the volume of the source improve the deviation of $R\out$, 
%but make the situation worse in $R\side$. 
%Contrary, some effects to increase the volume improve $R\side$, but make the situation worse in $R\out$. 
%We found that the effects of the mean field interaction is a candidate of such effects. 

\begin{figure}[t]
  \begin{center}  
  \includegraphics[scale=0.48]{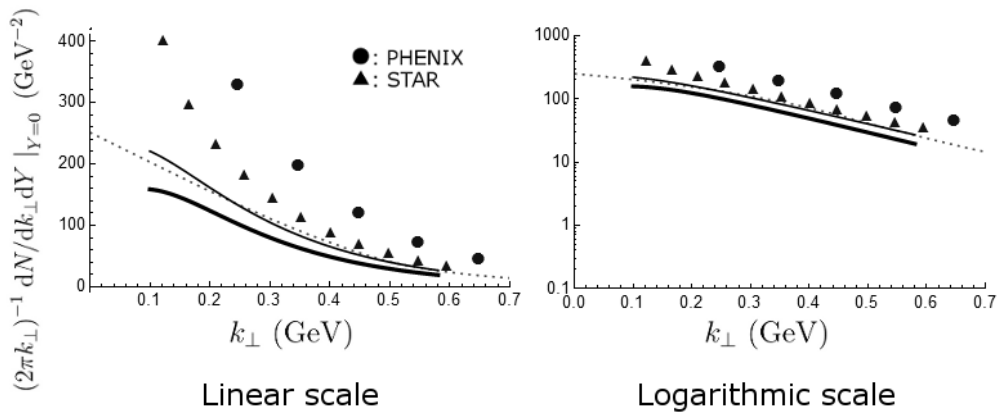}  
  \end{center}
\caption{
Modification of the transverse pion spectrum at the central rapidity: 
The dotted line indicates the free case, which is fixed on the freeze-out hypersurface, 
and the thin solid line the effects of the real part of the mean field potential. 
The thick solid line shows the effects of both the real and imaginary parts of the mean field potential. 
We show these effects in the linear and logarithmic scales, respectively. 
The circles and triangles exhibit the spectrum of the positive pion measured in 
Au+Au central collision at $\sqrt{s_{NN}}=200$ (GeV) in the RHIC experiment\cite{PtPHENIX,PtSTAR}. 
The negative pion is measured in almost the same amount. 
}
\label{fig:spec}
\end{figure}

For the sake of completeness, 
we show the modification of the transverse pion spectrum at the central rapidity. %by the mean field interaction. 
In Fig. \ref{fig:spec}, 
the dotted line displays the spectrum in the free case, which freezes abruptly at the hypersurface. 
The circles and triangles indicate the spectrum of the positive pion measured in 
Au+Au central collision at $\sqrt{s_{NN}}=200 \ {\rm GeV}$ in the RHIC experiment\cite{PtPHENIX,PtSTAR}. 
The negative pion is measured in almost the same amount. 
In the linear scale, we find that the effect of the attractive interaction acts to increase the distribution in the low-momentum regime, 
and to decrease that in the higher regime around $0.3$ to $0.4$ GeV. 
These effects show a deceleration of the emitted pions by the attractive mean field interaction. 
However, the effect of the real part is so slight that we find little deviation in the logarithmic scale. 
The distribution in the regime higher than $0.5$ GeV is barely affected by the real part of the mean field potential, 
because the shift of the momentum is much smaller than the magnitude of the momentum itself. 
The thick solid line indicates the effects of both the real and imaginary parts of the mean field potential.
The absorptive effect reduces the amount of pions below $0.6$ GeV, 
which is mainly due to the formation of $\rho$ meson resonance, $\pi\pi\rightarrow\rho$. 
Because the inverse reaction is not incorporated in our model as we have mentioned below Eq. (\ref{eq:reff}), 
the microscopic detailed balance is not maintained.\footnote{
Nonetheless, the normalization of the effective distribution function in Eq. (\ref{eq:reff}) is preserved 
even in the presence of the absorptive effect, as mentioned there. }
To refine the model, the finite lifetime of the resonance should be taken into account. 
Considering the decay process, $\rho\rightarrow\pi\pi$, which has the almost 100\% branching ratio, 
the pion number is conserved within the binary $\pi\pi$ collision.

Finally, we comment on the validity of the semiclassical method used in this analysis. 
If the classical action $S_c$ satisfies a condition 
\begin{eqnarray}
\left| \frac{\partial^2 S_c}{\partial \bm x^2} \right| \ll 
\left| \frac{\partial S_c}{\partial \bm x} \right|^2  \;,   \label{eq:cond}
\end{eqnarray}
we have Hamilton-Jacobi equation 
as the classical limit of the Schr\"odinger equation, 
and the approximation works well. 
The condition (\ref{eq:cond}) is rewritten as 
\begin{eqnarray}
\lambda = \frac{1}{\sqrt{2m(E-{\cal V}(\bm x))}} \ll 
2(E-{\cal V}(\bm x)) \left| \frac{d {\cal V}(\bm x)}{d \bm x} \right|^{-1} \;, \label{eq:cond1}
\end{eqnarray}
which indicates that the spatial profile of the potential ${\cal V}(\bm x)$ should not 
vary rapidly over a wavelength $\lambda$, 
and that the condition tends to be satisfied in the high-energy limit. 
Our computation is based on the semiclassical approximation applied to the four-dimensional pseudo-Schr\"odinger equation (\ref{eq:pSch}), 
in which ``energy", ``potential " and ``mass" are given by 
$\bar E = -m^2/2$, $\bar V(x^{\mu},\bm p)=\Pi(x^{\mu},\bm p)/2$ and $\bar m=1$, respectively. 
Thus, we have a condition to justify the approximation as an extension of Eq. (\ref{eq:cond1}). 
Using the physical quantities, the condition is represented by 
\begin{eqnarray}
\left( \frac{m_{\rm eff}}{m} \right)^{-3} \frac{1}{ 2 m^3 } 
\left| \frac{d \Pi(x^{\mu},\bm p)}{d x^{\mu}} \right| \ll 1 \;,  \label{eq:cond4}
\end{eqnarray}
where $m_{\rm eff}$ is the effective mass defined in Eq. (\ref{eq:meff}). 
%$m_{\rm eff}(x^{\mu},\bm p) = \sqrt{m^2 +\Pi(x^{\mu},\bm p)}$. 
%we have shown in Fig.\ref{fig:RePI}, 
%and the dimensionless quantity $m_{\rm eff}/m$ is in the order of one. 
Evaluating the left-hand side of Eq. (\ref{eq:cond4}), 
we obtain a sufficiently small value not larger than 0.05. 
Therefore, the semiclassical method employed in our analysis works well 
to evaluate the phase shift of the amplitude caused by the mean field interaction.

\section{Concluding remarks} \label{sec:CR}

In this work, we have examined the effects of a final state interaction on 
HBT interferometry in the ultrarelativistic heavy ion collision. 
This study is originally motivated by 
the so-called ``RHIC HBT puzzle", that is, an observation 
that the theoretical estimates of the HBT Gaussian radii show a systematic deviation 
from the experimental values measured in the RHIC experiments. 
To find a solution to this puzzle that has been addressed so far, 
we have studied how the mean field interaction 
ignored in the conventional hydrodynamical modeling 
may cause the distortion of the source images.

First, in \S \ref{sec:form}, 
we have examined how the effects of the mean field interaction are incorporated in the framework of HBT interferometry, 
beginning with the definition of the correlation function in terms of the density matrix. 
We found that the difference in the phase shifts imposed on each one-body amplitude of a pair 
induces the coordinate transformation in (\ref{eq:xshift}). 
It maps the original space-time coordinate onto the apparent one. 
The original grid, mapped onto the apparent coordinate system, 
is deformed owing to the nonlinear nature of the transform, as shown in Fig. \ref{fig:grid}. 
Then, the images viewed on the apparent coordinate system exhibit the distortions. 
%depending on the sign of the real part of the potential.  
We have shown that this is the geometrical interpretation of the distortion of the HBT images\cite{HM}.

In \S \ref{sec:prof}, 
we have shown that the images are distorted contrastingly depending on 
whether the interaction is repulsive or attractive: 
a repulsion acts to shrink the sideward extension and an attraction acts to stretch it, in contrast. 
The effect of the absorption cuts off the emissions in the interior and backside region of the source. 
This effect stretches the images in the sideward direction, 
which is somewhat similar to the effect of attractive interaction.

In \S \ref{subsec:meanfield}, a phenomenological model of the mean field interaction is constructed 
on the basis of the elastic forward $\pi\pi$ scattering amplitude and 
the pion distribution function after the ``freeze-out". %within the free streaming. 
Owing to the dominant contribution of the p-wave scattering, 
we found the attractive mean field interaction in the low-momentum regime in some hundreds MeV. 
In this regime, the mean field interaction acts on the images effectively. 
In the high-momentum regime, 
the attraction turns into repulsion due to the repulsive regime 
in the p-wave scattering beyond the $\rho$ meson mass. 
However, we did not find a considerable effect of the repulsive interaction. %on the Gaussian radius parameters. 

Using the phenomenological self-energy obtained in \S \ref{subsec:meanfield}, 
we have computed the modification of the amplitude of an emitted pion, 
that is, the phase shift and attenuation factor. 
We have employed the semiclassical method given in Appendix A. 
We examined how and to what extent the modification results in the Gaussian radius parameters. 
We found that the effect of the obtained attractive interaction actually increases the sideward radius $R\side$ 
as implied in the preceding works\cite{CMWY,Pra06,HM}, 
and that it improves the deviations in $R\side$ and the ratio, $R\out/R\side$. 
However, the magnitude of the effect is not sufficiently strong to resolve the deviations completely, 
since the pion density is not sufficiently large in the exterior of the freeze-out hypersurface 
owing to the evaporation. 
%Especially, when the emission on the transverse plane occurs in the surface region 
%in advance to the interior, pions sequentially escape from the vicinity of the source. 
%Therefore, the density of the emitted pion is likely to be low 
%compared to the case in the instantaneous emission, or the static case, 
%which prevents the emitted pions from forming a thick cloud. 
%A consistent description of the dynamics with the mean field interaction 
%is necessary to investigate its effects throughout the hadron phase. 
%This time dependence of the freeze-out hypersurface prevents pions from forming a thick cloud. 
%We conclude that 
%the effect of the mean field interaction, in the exterior of the freeze-out hypersurface, 
%improves the deviations in $R\side$ and the ratio, $R\out/R\side$, 
%however, the resolution of the problem is not provided 
%with the sole effect of the mean field interaction by the pion cloud. 
%The repulsive interaction obtained in the higher momentum regime 
%does not act on the Gaussian parameters effectively, 
%since the effects of the mean field interaction diminishes as the momentum gets larger. 

We briefly comment on the related works. 
It has recently been suggested that 
some upgrades of the hydrodynamical model lead to 
considerable improvements of the deviations owing to the cooperative effects of the upgrades\cite{Pra09}. 
We note that the modification of $R\side$ obtained in this work 
is on the same order of their individual effects. 
The mean field interaction in the matter, not included in our model, would act to vary the apparent radii more efficiently, %provide more strong effects, 
since the density is larger than that in the very last stage. 
%While it may be redundant for the hydrodynamical modeling to incorporate the effect of the interaction with the matter, further investigation is required to 
%it should be incorporated in the kinetic equations via the Vlasov term. 
This possibility should be examined 
with the consistent description of the mean field interaction throughout the hadron phase\cite{MM}. 
%Carrying out the simulations incorporating the Vlasov term and collision term into the transport equation, 
%the self-energy would be treated self-consistently 
%with respect to the motion of medium particle, as is mentioned in \S \ref{subsec:meanfield}. 
Alternatively, if we attempt to treat the effects of the mean field interaction in the hydrodynamic picture, 
we could study it with the equation of state. %incorporating its effect. 
Its effects possibly reflect in the profile of the freeze-out hypersurface.

\section*{Acknowledgments}

This work is partially based on the author's thesis submitted to 
the Department of Physics, The University of Tokyo. 
The author is grateful to his thesis supervisor Prof. T. Matsui for 
the instruction and encouragement. 
He would like to thank Prof. T. Matsui for the comments on the composition of the manuscript, 
Prof. H. Fujii for the useful advices and careful reading of his thesis, 
and Prof. T. Hirano for the discussion in our early work. 
He also thanks Prof. O. Morimatsu and Prof. K. Itakura for the helpful comments, 
and Prof. T. Hatsuda, Prof. T. Otsuka, Prof. K. Ozawa and Prof. H. Sakurai for the discussions. 
He sincerely thanks Prof. K. Yazaki for the conversations on the related works, 
and Prof. J.-P. Blaizot for the useful comments. 
This work was supported in part by the Global COE Program ``the Physical Sciences Frontier", MEXT, Japan.

\section*{Appendix A} \label{sec:appA}

We examine the modification of the amplitude incorporating the relativity. 
%still adopting the semi-classical approximation. 
The amplitude of the pseudo scalar particle obeys the Klein-Gordon equation 
\begin{eqnarray}
\left( \Box + m^2 + \Pi(x) \right) \varphi_{\bm k}(x)= 0 \ ,  \label{eq:KGeq}
\end{eqnarray}
where the coordinate $x$ denotes the four-vector $x^{\mu}=(t,\bm x)$, 
and the box stands for the derivative operator, $\Box = \partial_t^2 - \partial_{\bm x}^2$. 
The effect of the mean field interaction is incorporated in the self-energy $\Pi(x)$, 
which is the mass shift of the pion mass, $m$, in the vacuum. 
The modification of the dispersion relation in the pion cloud 
is phenomenologically studied in \S \ref{subsec:meanfield}. 
In this appendix, we examine a method of obtaining the semiclassical form of the amplitude.

A parameter called the proper time enables us to adopt the semiclassical approximation 
to the Klein-Gordon equation (\ref{eq:KGeq}) 
employing the standard prescription in quantum mechanics\cite{Fey50,Sch}. 
Using the proper time $\lambda$ , the Klein-Gordon equation (\ref{eq:KGeq}) reads 
\begin{eqnarray}
\left\{
\begin{array}{l}
\left( \frac{\Box}{2}  + \frac{\Pi(x)}{2} \right) \Psi_{\bm k}(x,\lambda)
= i \frac{\partial }{\partial \lambda} \Psi_{\bm k}(x,\lambda) \ , \\
\Psi_{\bm k}(x,\lambda) = e^{i\frac{m^2}{2} \lambda} \varphi_{\bm k}(x) \ ,
\end{array}
\right. 
\label{eq:pSch}
\end{eqnarray}
where $\Psi_{\bm k}(x,\lambda)$ is the amplitude 
that obeys the pseudo-Schr\"odinger equation (\ref{eq:pSch}) with the unit mass, 
and the pseudo potential is given by the self-energy, $\Pi/2$. 
The parameter $\lambda$ plays a role of ``time" in nonrelativistic quantum mechanics, 
and $\Psi_{\bm k}(x,\lambda)$ is the eigenfunction of ``energy" given by $-m^2/2$. 
We deduce that 
the pseudo-Schr\"odinger equation (\ref{eq:pSch}) is obtained from the Hamiltonian in the four dimensions, 
\begin{eqnarray}
\bar H(\hat x, \hat p) = -  \frac{\hat p^2}{2} + \frac{\Pi(\hat x)}{2} \;, \label{eq:hamilton4}
\end{eqnarray}
where we take a convention of the Minkowski metric, $\hat p^2 = \hat p_0^2 - \hat {\bm p}^2$.

Owing to the argument in \S \ref{subsec:form2}, 
the transition amplitude $\varphi( \bm x^\prime,t^\prime; \bm x, t )$ 
from $(t,\bm x)$ to $(t^\prime,\bm x^\prime)$ provides 
the amplitude $\varphi_{\bm k}(x)$ of a particle emitted at $(t,\bm x)$ 
to detect with the asymptotic momentum $\bm k$, %up to the semi-classical approximation, 
as long as the classical trajectory is uniquely specified by $\bm x$ and $\bm k$. 
Assuming that the transition amplitude $\varphi(\bm x^\prime,t^\prime; \bm x, t)$ 
obeys the Klein-Gordon equation (\ref{eq:KGeq}), 
the technique with the proper time enables us to write 
\begin{eqnarray}
\varphi(\bm x^\prime,t^\prime; \bm x, t) = 
e^{-i\frac{m^2}{2} \lambda} \langle x^{\prime} | e^{-i \bar H \lambda} | x \rangle \;. \label{eq:3to4}
\end{eqnarray}
On the right-hand side, the transition amplitude, from $x^{\mu}$ to $x^{\prime\mu}$ 
in the ``time" interval $\lambda$, obeys the pseudo-Schr\"odinger equation (\ref{eq:pSch}). 
Using the conventional prescription, 
we obtain the path integral form of the amplitude: 
\begin{eqnarray}
\langle x^{\prime} | e^{-i \bar H \lambda} | x \rangle = 
\int \!\! \mathscr{D}x \! \int \!\! \mathscr{D}p \:
e^{ i \int_0^{\lambda} d\lambda^\prime \ \left( - p^{\mu} \acute x_{\mu} - \bar H(x,p) \right) } \ ,
\label{eq:Rsemi}
\end{eqnarray}
where the acute on the coordinate denotes the derivative with respect to the proper time as 
$$ - p^{\mu} \acute x_{\mu} = - p^0 \frac{d t}{d\lambda} 
+ \bm p \cdot \frac{d \bm x}{d\lambda} \:\:\:.$$

We evaluate the right-hand side of Eq. (\ref{eq:Rsemi}) adopting the stationary phase approximation. 
The variation of the real part, $\delta(  p^{\mu} \dot x_{\mu} + {\rm Re}\, \bar H(x,p) )$, 
leads to the canonical equations of motion 
\begin{eqnarray}
\left\{
\begin{array}{l}
\acute x^{\mu} = - \frac{ \partial \bar H_r}{ \partial p_{\mu}}  \ , \\
\acute p^{\mu} = \frac{ \partial \bar H_r}{ \partial x_{\mu}} \ , 
\end{array}
\right. \label{eq:cano4}
\end{eqnarray}
where $\bar H_r$ is the real part of the Hamiltonian, $\bar H_r(x,p)={\rm Re} \,\bar H(x,p)$. 
The proper time parameterizes the world line of the classical motion. 
Using the temporal component of Eq. (\ref{eq:cano4}), 
we eliminate the proper time in the spatial components 
and parametrize the world line using the physical time, $t$. 
%: this is the conventional choice of the parametrization as we  to write down the one-particle classical action 
%in relativistic mechanics. 
Substituting the Hamiltonian (\ref{eq:hamilton4}), 
the spatial components of Eq. (\ref{eq:cano4}) are expressed as 
\if 0
\begin{eqnarray}
\left\{
\begin{array}{l}
\dot {\bm x} = - \frac{ \partial \bar H}{ \partial \bm p} / \frac{ \partial \bar H}{ \partial p^0}  \ , \\
\dot {\bm p} = - \frac{ \partial \bar H}{ \partial \bm x} / \frac{ \partial \bar H}{ \partial x^0} \ , 
\end{array}
\right. \label{eq:cano}
\end{eqnarray}
\fi
\begin{eqnarray}
\left\{
\begin{array}{l}
\dot {\bm x} = \frac{ \bm p }{ E_{\bm p} }  \ , \\
\dot {\bm p} = - \frac{1}{2 E_{\bm p}} 
\frac{ \partial \Pi_r(\bm x,t)}{ \partial \bm x} \ , 
\end{array}
\right. \label{eq:cano}
\end{eqnarray}
where $\Pi_r(x)$ is the real part of the self-energy, $\Pi_r(x)={\rm Re}\,\Pi(x)$, 
and the dot indicates the derivative with respect to time, $t$. 
The zeroth component of $p^{\mu}$ is given by $E_{\bm p}=\sqrt{\bm p^2 + m^2 + \Pi_r}$. 
The set of canonical equations (\ref{eq:cano}) determines the classical trajectory 
under the existence of the mean field interaction.

During the classical motion,  the ``energy" given by $-m^2/2$ is conserved, 
because the Hamiltonian (\ref{eq:hamilton4}) does not depend on the proper time. 
Then, we have a dispersion relation, 
$$ \bar H_r(x,p) = - \frac{p^2}{2} + \frac{\Pi_r(x)}{2}  = -\frac{m^2}{2} \ ,$$ 
where the Hamiltonian is regarded as {\it c}-number under the semiclassical approximation. 
Substituting the above expression and Eq. (\ref{eq:Rsemi}) for Eq. (\ref{eq:3to4}), 
we obtain the transition amplitude 
\begin{eqnarray}
\varphi(\bm x^\prime,t^\prime; \bm x, t) = 
\mathscr{A} e^{ i \left( \integ \bm p \cdot d\bm x - \integ E_{\bm p} dt \right) }  \ , \label{eq:ramp0}
\end{eqnarray}
where the integrals are performed along the world line determined with Eq. (\ref{eq:cano}). 
Owing to the change in the integral variable using Eq. (\ref{eq:cano}), 
we find a simple extension to the relativistic form, 
\begin{eqnarray}
\varphi(\bm x^\prime,t^\prime; \bm x, t) &=& \mathscr{A}  e^{iS(\bm x^\prime,t^\prime; \bm x, t)} \ , \nonumber\\
S(\bm x^\prime,t^\prime; \bm x, t) &=& 
- \int_t^{t^\prime} m_{{\rm eff}} \sqrt{ 1 - \dot{\bm x}^2 } \ dt^{\prime} \ .
\label{eq:ramp}
\end{eqnarray}
The effect of the mean field interaction is incorporated in the effective mass, 
$m_{{\rm eff}}  = \sqrt{m^2+ \Pi_r(x)}$, 
appearing in the relativistic one-particle action $S(\bm x^\prime,t^\prime; \bm x, t)$. 
Taking the imaginary part of the Hamiltonian in Eq. (\ref{eq:Rsemi}), 
the prefactor $\mathscr{A}$ associated with the classical trajectory 
is provided by integrating the imaginary part of the self-energy along the world line, 
\begin{eqnarray}
\mathscr{A} = %a(\bm x, 0; \bm x^\prime,t) 
\ \exp {\int \!\! \frac{1}{ 2 E_{\bm p} } {\rm Im} \left[ \Pi(x) \right] dt} \ ,
\label{eq:att}
\end{eqnarray}
where we have used the temporal component of Eq. (\ref{eq:cano4}) 
to take time, $t$, as the integral variable.

As mentioned above, the semiclassical evaluation of $\varphi(x^{\prime}, x)$ 
provides the desired amplitude $\varphi_{\bm k} (x)$. 
As the four-dimensional analogue of Eq. (\ref{eq:kx}), 
we have 
\begin{eqnarray} 
\varphi_{\bm k} (x) = \integ dx^\prime \  e^{ikx^\prime} \  \varphi(x^{\prime}, x) \ ,
\label{eq:1}
\end{eqnarray}
where $x$ is the coordinate four vector, 
and $k$ is the asymptotic momentum that satisfies the on-shell condition, 
$E_\bk^2=\bk^2+m^2$. 
If we choose sufficiently a large time $t_1$ and 
spatial coordinate $\bm x_1$ far outside the range of the mean field potential, 
the relativistic classical action in Eq. (\ref{eq:ramp}) is decomposed into two parts 
as in Eq. (\ref{eq:SS0}), 
\begin{eqnarray}
S(x^{\prime}, x) &=& S(x^{\prime}, x_1) + S(x_1,x)  \ , \nonumber\\
&=&  k (x^{\prime}-x_1) + S(x_1,x) \:\:, \nonumber
\end{eqnarray}
where $S(x^{\prime},x_1)$ is the classical action of the free motion from $x_1$ to $x^{\prime}$. 
%, and the momentum $k_0$ is given by a constant vector. 
Inserting the above relation into Eq. (\ref{eq:1}) and performing the integral, 
the amplitude is given by 
\begin{eqnarray} 
\varphi_{\bm k} (x) &=& \mathscr{A}\  e^{-i kx + i \delta S(x_1,x)} \label{eq:rela1}   \\
\delta S(x_1,x) &=&  S(x_1,x) - k(x_1-x) 
\ , \label{eq:rela2}
\end{eqnarray}
where $\delta S(x_1,x)$ is the phase shift caused in the range of the mean field potential. 
%, and the zero-th component of $k$ is given by $k_0=\sqrt{\bm k^2+m^2}$. 
Using the formulae in Eqs. (\ref{eq:rela1}) and (\ref{eq:rela2}), 
we obtain the amplitude from the computation of the relativistic one-particle action, 
the integral of which is performed along the world line determined using the canonical equations (\ref{eq:cano}).

We remark that the dynamical mean field potential, 
or phenomenological self-energy, studied in \S \ref{subsec:meanfield} 
depends on the momentum as well as the space-time coordinate. 
Then, the canonical momentum appearing in the Hamiltonian %and eq.(\ref{eq:Rsemi}) 
should be distinguished from the kinetic momentum defined by the velocity. 
However, we neglect their difference which is given by the derivative of the self-energy with respect to the momentum, 
since it is negligibly small in our model compared with the magnitude of the momentum in some hundreds of MeV.

\end{document}